\documentclass[a4paper,11pt]{article}
\pdfoutput=1
\usepackage{jheppub}

\usepackage[T1]{fontenc}
\usepackage{amssymb}
\usepackage{amsmath}
\usepackage{graphicx}
\usepackage{subcaption}
\usepackage[colorlinks=true]{hyperref}
\usepackage{appendix}
\usepackage{amsbsy}

\usepackage{booktabs}
\usepackage{multirow}

\usepackage{tabularx}
\usepackage{pifont}

\def\beq{\begin{equation}}   
\def\eeq{\end{equation}}
\def\bea{\begin{eqnarray}}  
\def\eea{\end{eqnarray}} 
\def\nn{\nonumber}

\def\doubletilde#1{\widetilde{\vphantom{\raise 1.5pt \hbox{#1}}\smash{\kern -2pt\widetilde{#1}}}}

\def\e{\epsilon}
\def\d{\hbox{d}}

\def\nn{\nonumber}

\def\e{\epsilon}
\def\d{\hbox{d}}

\def\nn{\nonumber}

\newcommand{\qb}{{\bar{q}}}

\newcommand{\wt}[1]{\widetilde{#1}}

\newcommand\pll{\parallel}

\newcommand\soft[1]{\mathbb{S}\left({#1}\right)}
\newcommand\coll[1]{\mathbb{C}\left({#1}\right)}

\newcommand{\lb}{\left\lbrace}
\newcommand{\rb}{\right\rbrace}
\newcommand{\vardbtilde}[1]{\tilde{\raisebox{0pt}[0.85\height]{$\tilde{#1}$}}}

\usepackage{xspace}
\usepackage[dvipsnames]{xcolor}

\preprint{{\raggedleft%
IPPP/25/44
}}

\title{Phase-space sectors for ordered momentum mappings in local subtraction up to N$^3$LO}

\author[a]{Xuan Chen,}
\author[b]{Matteo Marcoli}

\affiliation[a]{School of Physics, Shandong University,
            Jinan, Shandong 250100, China}
\affiliation[b]{Institute for Particle Physics Phenomenology, Physics Department, Durham University, Durham, DH1 3LE, UK}
\emailAdd{xuan.chen@sdu.edu.cn}
\emailAdd{matteo.marcoli@durham.ac.uk}

\abstract{%
	Ordered momentum mappings present optimal convergence in soft and collinear configurations and are particularly suitable for the numerical implementation of local subtraction schemes.  However, ordered mappings cannot be directly applied in the presence of multiple unordered emissions, which typically appear beyond the leading-colour approximation. A possible solution consists in separating individual singularities at the level of local counterterms by means of partial fractioning, which can become cumbersome at higher orders and introduce large cancellations in intermediate steps of the calculations. We present a simple decomposition of the phase space into sectors to isolate classes of infrared configurations which can be addressed with a specific ordered momentum mapping. With such decomposition, the singularities of any matrix element can be subtracted with ordered mappings, without the need of partial fractioning. We illustrate the required phase-space sectors for up to three unordered emissions and discuss applications in the context of the antenna subtraction method. The mapping algorithm described here has been recently employed for the  first differential N$^3$LO of jet production at electron-positron colliders.
}

\begin{document} 
\maketitle
\flushbottom

\section{Introduction}\label{sec:intro}

The increasing precision of measurements performed at the LHC and the projected accuracy for future collider experiments call for equally precise theoretical predictions. Calculations in perturbative QCD enter the predictions for the vast majority of high-energy collider observables and are therefore central to improve the overall accuracy of the theoretical description. The major complication in computing perturbative corrections on top of the Leading Order (LO) prediction consists in the appearance of infrared singularities in both virtual and real corrections. These arise when massless propagators in tree-level or loop matrix elements approach the on-shell limit and become manifest as explicit poles in dimensional regularisation or as implicit divergences when external particles are emitted in soft and collinear configurations. Such infinities cancel in final predictions for sufficiently inclusive observables, but need to be properly treated to perform fully differential calculations for arbitrary scattering processes. 

Infrared subtraction methods are algorithms designed to capture infrared singularities and achieve their (guaranteed) cancellation while also enforcing that contributions to physical cross sections coming from different partonic multiplicities are separately finite and computable with numerical techniques.  The first perturbative correction, namely the Next-to-Leading Order (NLO), has been under control for decades, thanks to the availability of general techniques to compute it~\cite{Catani:1996jh,Frixione:1995ms}. Next-to-Next-to-Leading Order (NNLO) calculations pose a much greater challenge, because of the complexity of the necessary infrared subtraction, but also due to the higher computational cost. Several approaches have been formulated~\cite{Gehrmann-DeRidder:2005btv,Boughezal:2011jf,Currie:2013vh,DelDuca:2016ily,Catani:2007vq,Czakon:2010td,Czakon:2014oma,Gaunt:2015pea,Cacciari:2015jma,Caola:2017dug,Magnea:2018hab,Herzog:2018ily,TorresBobadilla:2020ekr,Bertolotti:2022aih}, with increasing recent efforts to design more general and more efficient techniques~\cite{Gehrmann:2023dxm,Fox:2024bfp,Bonino:2024adk,Devoto:2023rpv,VanThurenhout:2024hmd,Devoto:2025kin,Bertolotti:2025clg}. Various computations of the NNLO correction to high-multiplicity $2\to 3$ processes~\cite{Chawdhry:2019bji,Kallweit:2020gcp,Chawdhry:2021hkp,Czakon:2021mjy,Chen:2022ktf,Hartanto:2022qhh,Alvarez:2023fhi,Badger:2023mgf,Catani:2022mfv,Buonocore:2022pqq,Buonocore:2023ljm,Mazzitelli:2024ura,Devoto:2024nhl,Buccioni:2025bkl} showcase the level of maturity reached by NNLO subtraction schemes.

The mechanism at the core of subtraction schemes, and in particular of \textit{local} subtraction schemes, is the transferring and cancellation of infrared singularities between phase spaces of different dimensions. Generally, \textit{subtraction terms} are built exploiting the universal factorisation properties of QCD matrix elements to remove their divergent behaviour in soft and collinear limits. This requires mapping a generic point of an $(n+k)$-dimensional phase space, where $k$ momenta goes unresolved, to a hard configuration in a $n$-dimensional phase space. The divergent subtraction terms are then integrated in dimensional regularisation exploiting the phase-space factorisation induced by the mapping and added back in the phase space with matching dimension where they cancel the explicit singularities coming from loop integrals. 

A local subtraction term for the removal of the divergent behaviour of an $(n+k)$-parton real-emission matrix element $M_{n+k}(\{p\}_{n+k})$, with $\{p\}_{n+k}$ being the set of parton momenta, can generically be written as:
\begin{equation}\label{eq:subterm}
	\d\sigma^{S}=X(\{p\}_{k})\otimes M_{n}(\{P\}_{n}) J(\{P\}_{n}),
\end{equation}
where the set of hard momenta $\{P\}_{n}$ is obtained via a suitable map $\{p\}_{n+k}\to\{P\}_{n}$ to reabsorb the recoil of the $k$ unresolved emissions. $X(\{p\}_{k})$ is a function of the unresolved momenta and potentially hard spectators (which are omitted here for simplicity) encoding the infrared singularities, while $M_{n}(\{P\}_{n})$ and $J(\{P\}_{n})$ are the Born-level matrix element and the measurement function implementing the definition of the observables. The symbol $\otimes$ is a shorthand notation to indicate that in general unresolved factors are operators acting on colour- and spin-correlations of the Born-level matrix element. Moreover, it also represents the fact that the $\{p\}_{n+k}\to\{P\}_{n}$ map is induced by the unresolved factor. Indeed, for the cancellation of divergences to occur, the mapped hard momenta $\{P\}_{n}$ have to be properly reconstructed according to the considered infrared configuration. 

The unresolved factors and associated momentum mappings are the core ingredients for a subtraction method. A common approach to systematically build~\eqref{eq:subterm} in local subtraction schemes consists in partitioning the phase space into sectors where only individual infrared configurations are allowed~\cite{Frixione:1995ms,Czakon:2010td,Caola:2017dug,Magnea:2018hab}. In each sector, the divergent behaviour of real-emission matrix elements, hence the choice of unresolved factor $X(\{p\}_{k})$, is uniquely fixed by the factorization properties of QCD in the selected infrared limit. The form of the momentum mapping also follows from the phase-space restriction. In particular, the mapping has to yield the proper hard momenta only in the selected limit and is not required to have any specific behaviour elsewhere. The main challenge when following this approach is the identification of a consistent partition of the phase space to isolate each individual limit, which is highly non-trivial in the presence of multiple unresolved emissions. An alternative strategy consists in adopting unresolved factors which can interpolate between different infrared configurations by reproducing the correct divergent behaviour of real-emission matrix elements in multiple limits~\cite{Catani:1996vz,Gehrmann-DeRidder:2005btv,DelDuca:2016ily}. This prevents the need of partitioning the phase-space, but poses other challenges. One of them is the choice of the momentum mapping, which is now required to yield the appropriate hard momenta in multiple unresolved limits and is therefore significantly more constrained. 

The antenna subtraction scheme~\cite{Gehrmann-DeRidder:2005btv,Currie:2013vh}, implemented in the Monte Carlo framework NNLOJET~\cite{NNLOJET:2025rno} is an example of the latter approach. It has been successfully applied for the computation of the NNLO corrections to several processes, in particular involving final-state jets. It is based on \textit{antenna functions} which, traditionally, have been extracted from physical colour-ordered squared amplitudes and  capture the divergent behaviour of multiple unresolved configurations arising from  soft and/or collinear emissions from two hard radiators~\cite{Gehrmann-DeRidder:2004ttg,Gehrmann-DeRidder:2005alt,Gehrmann-DeRidder:2005svg}. Recently, an algorithm has been formulated to build antenna functions directly from unresolved factors~\cite{Braun-White:2023sgd,Braun-White:2023zwd,Fox:2023bma} and go beyond the two-hard-radiators paradigm~\cite{Fox:2024bfp}. With antenna functions as designated unresolved factors, the momentum mapping needs to handle several infrared configurations. A natural choice of momentum mapping for final-state radiation in two-hard-radiator antenna functions is the so-called antenna mapping~\cite{Kosower:1997zr,Kosower:2002su}. It absorbs an arbitrary number of emissions into two hard momenta, preserving momentum conservation and on-shellness conditions. Its advantage lies in the capability of smoothly transitioning between multiple unresolved configurations. However, despite its versatility, there is still a limitation to the possible unresolved configurations it can simultaneously handle. Indeed, the antenna mapping is an \textit{ordered mapping}: for a given ordering of the unresolved emissions \mbox{$\{p_a^h,p_1,...,p_n,p_b^h\}$}, where the superscript $h$ indicates the two hard momenta, the antenna mapping \mbox{$\{p_a^h,p_1,...,p_n,p_b^h\}\to \{P_a,P_b\}$} reconstructs the correct two-hard-particle configuration given by the momenta $P_a$ and $P_b$ only in those infrared limits which are allowed by the ordering. More specifically, if an unresolved limit involves collinear momenta which are not adjacent in the ordering, the antenna mapping fails to reconstruct the hard momenta. Since antenna functions are designed to target singularities of colour-ordered squared amplitudes, the ordering property of the antenna mapping does not in general pose a problem. However, beyond the leading-colour approximation, unordered emissions can occur, for example when multiple abelian gluons, namely gluons which have collinear limits with non-adjacent partons in the colour string, hence behaving like photons, are emitted. In such a case, both the real-emission matrix elements and the antenna functions exhibit a divergent behaviour regardless of the ordering of the momenta, and thus the antenna mapping cannot be straightforwardly employed. 

A possible solution to this problem consists in modifying the unresolved factors, namely the antenna functions, to split up the infrared limits into pieces to which a different momentum mapping can be assigned. This approach has been applied in the context of antenna subtraction up to NNLO: problematic antenna functions describing multiple unordered emissions are split into so-called \textit{sub-antennae}~\cite{Gehrmann-DeRidder:2007foh,Pires:2010jv}. These are obtained from full antenna functions by partial fractioning, isolating the vanishing denominators in specific infrared limits, in such a way that each sub-antenna contains the singularities of a specific ordering. Although it works well at NNLO, this strategy has critical drawbacks which shall become important at N$^3$LO and beyond. First of all, sub-antennae are typically more complicated than the original antenna functions because repeated partial fractioning leads to a proliferation of terms. The number of orderings and hence of singular denominators which need to be separated grows factorially with any additional emission, therefore multivariate partial fractioning of unresolved factors can become a serious bottleneck with three or more emissions. Moreover, even if one is careful about avoiding the appearance of spurious denominators, when an antenna function is split into sub-antennae, singular terms which do not need to be isolated, for example the soft divergences, get split into multiple contributions, potentially resulting in large cancellations in intermediate steps of the calculation. Finally, from N$^3$LO on, unresolved factors targeting multiple emissions in loop matrix elements need to be considered, which usually feature non-rational singular contributions, for which manipulations to separate individual limits would be highly non-trivial. 

In this paper, we present an alternative solution, which prevents the need of partial fractioning, easily scales with the number of unresolved emissions and is independent from the specific expressions of the antenna functions. The idea consists in keeping the unresolved factors unchanged and instead acting on how the momentum mapping is assigned to them. In particular, for any unresolved emission multiplicity up to N$^3$LO, we adopt the well-established strategy of separating the phase space into sectors where only a given subset of infrared configurations can exist to discriminate between different choices of ordered mapping for a given antenna function. Since we do not need to isolate each individual infrared limit, but rather just classes of limits requiring a specific ordering in the antenna mapping, the phase-space partition we need to define for this purpose is quite simple.

The illustration of the approach we present is carried on in the context of the antenna subtraction method, for the sake of providing explicit examples of the problematic configurations, comparing with alternative solutions and presenting validation at the numerical level. However, the underlying implementation is more general and can in principle be applied in other contexts. The paper is structured as follows. In Section~\ref{sec:basics} we illustrate our notation and review the basics of the antenna subtraction method. In Section~\ref{sec:sectors} we present the definition of the phase-space sectors at NNLO and N$^3$LO and  in Section~\ref{sec:validation} we provide analytical and numerical evidence supporting the correctness of the algorithm. Finally, in Section~\ref{sec:conclusions} we draw our conclusions. 


\section{Notation and basics of antenna subtraction}\label{sec:basics}

In this section we introduce our notation and briefly review the concepts of antenna subtraction which are relevant for the reminder of the paper.

\subsection{Infrared limits}

When infrared limits with multiple unresolved partons are considered, several configurations of soft and collinear momenta are possible. We introduce the following shorthand symbols to describe a given infrared phase-space region:
\begin{itemize}
	\item $\soft{1,\dots,n}$: $p_i\to 0$, $i=1,\dots,n$, to indicate momenta becoming soft,
	\item $\coll{1,\dots,n}$: $p_1\pll \dots \pll p_n$, to indicate momenta becoming collinear to each other,
\end{itemize}
which can be combined with the $\otimes$ symbol to indicate composite unresolved configurations. For example, the triple-unresolved limit $p_a\to 0,\, p_b\to0,\, p_c\pll p_d$ will be indicated as $\soft{a,b}\otimes\coll{c,d}$. We observe that the notation above does not assume any particular behaviour of the matrix elements in the considered limits, but simply refers to the behaviour of the particle momenta. For example, in the region $\soft{a,b}$, a given colour-ordered matrix element may exhibit a double-soft divergence or an iterated single-soft divergence according to the colour connections it contains.

\subsection{Antenna functions}

Antenna functions encapsulate the divergent behaviour of matrix elements in infrared limits~\cite{Gehrmann-DeRidder:2005alt} and are fundamental ingredients used in the antenna subtraction method. A general $n$-particle, $\ell$-loop antenna function is denoted with:
\begin{equation}
	X_{n}^{\ell}(i_1,\dots,i_n),
\end{equation}
where $i_1,\dots,i_n$ label the partons. Different letters can be used in place of $X$ to indicate specific partonic configurations. A hat ($\hat{X}$) indicates the presence of a closed fermionic loop, while a \text{tilde} ($\tilde{X}$) indicates subleading-colour contributions.  The number $n$ represents the sum of hard radiators and unresolved partons. In the following, we will only consider antenna functions with two hard radiators, because these are the ones affected by the results we present.  Generalised antenna functions are briefly discussed in Section~\ref{sec:du_map}.

Antenna functions can be analytically integrated over the inclusive phase space of the associated unresolved radiation. A proper definition of the antenna phase-space is given later in Section~\ref{sec:mapping}. For final-state radiation, two-hard-radiator integrated antenna functions are generically indicated as
\begin{equation}
	\mathcal{X}_{n}^{\ell}(s_{i_1\dots i_n})=\int\text{d}\Phi_{n}\,X_{n}^{\ell}(i_1,\dots,i_n)
\end{equation}
where $s_{i_q\dots i_n}$ is the invariant mass of the considered partonic system:
\begin{equation}
	s_{i_1\dots i_n}=\left(\sum_{i=1}^n p_i\right)^2.
\end{equation}
To perform an N$^k$LO calculation, one needs $X_n^\ell$ and $\mathcal{X}_n^\ell$ for all the ($n$,\,$\ell$) pairs which satisfy $(n-2)+\ell\leq k$, with $n>2$.

Antenna functions are traditionally extracted from physical matrix elements~\cite{Gehrmann-DeRidder:2004ttg,Gehrmann-DeRidder:2005alt,Gehrmann-DeRidder:2005svg,Jakubcik:2022zdi,Chen:2023fba,Chen:2023egx} and they typically contain all the unresolved limits associated to a specific partonic configuration.  On one hand this allows them to interpolate across several infrared limits without the need to restrict their application in corners of the phase space. On the other, in some cases it is not possible to assign to individual antenna functions a momentum mapping, namely a prescription describing how to reabsorb additional emissions into the underlying hard event, which preserves all the associated unresolved configurations. We can distinguish two scenarios where this is not possible:
\begin{enumerate}
	\item an antenna function contains infrared limits which correspond to different choices of hard radiators;
	\item an antenna function contains multiple unordered emissions;
\end{enumerate}
The common strategy employed so far to deal with these complications consisted in splitting up the antenna functions into pieces which only contain a subset of the original singularities. We discuss this in more detail in Section~\ref{sec:subantenna}, where we also provide explicit examples of the issues above.

The first problematic scenario above has been solved in the general case by the recently proposed \textit{designer antenna algorithm}~\cite{Braun-White:2023sgd,Braun-White:2023zwd,Fox:2023bma}. In this context, antenna functions are not extracted from physical matrix elements, but directly assembled (\textit{designed}) from individual singular factors according to the desired unresolved limits. The antenna functions obtained with this strategy benefit from the significant advantage of having a well-defined pair of hard radiators, so that the ambiguity of the hard radiator choice disappears from the assignment of the momentum mapping. It is foreseeable that extending the algorithm at N$^3$LO will allow one to construct antenna functions with well defined hard radiators at this perturbative order too. The designer antenna functions, also called \textit{idealised antenna functions}, however, still suffer from the second issue above, because the presence of singular denominators belonging to different partonic orderings is an intrinsic feature of the unresolved factors for multiple unordered emissions.

\subsection{Antenna phase-space factorisation and mappings up to N$^3$LO}\label{sec:mapping}

We review here the factorisation of the antenna phase-space in the case of final-state radiation and the choice of momentum mapping which is used in the implementation of antenna subtraction.

The Lorentz invariant phase space $\d \Phi_n$ for $n$ particle in the final state, each with momentum $p_i$, in $d=4-2\epsilon$ space-time dimensions reads
\begin{equation}
	\label{eq:PSdefinition}
	\d \Phi_n (p_1,\cdots,p_n;q)=\frac{\d^{d-1} p_1}{2E_1(2\pi)^{d-1}}\dots \frac{\d^{d-1} p_n}{2E_n(2\pi)^{d-1}}(2\pi)^d\delta^d(q-p_1-p_2\cdots-p_n).
\end{equation}
The antenna phase-space for an $m$-particle antenna function is defined as~\cite{Gehrmann-DeRidder:2005btv}:
\begin{equation}\label{eq:antPS}
	\d \Phi_{X_m}(p_1,\dots,p_m;P_1+P_2) = \dfrac{\d \Phi_m(p_1,\cdots,p_m;q)}{\int \d \Phi_2(P_1,P_2;q)},
\end{equation}
where $P_1$ and $P_2$ are obtained from the momenta $\lb p \rb_n$ via a $m\to 2$ mapping which preserves momentum conservation ($P_1+P_2=q$) and on-shellness conditions. The explicit form of the mapping is not relevant to computing the denominator of~\eqref{eq:antPS}, which is given by the inclusive integration over the two-particle phase space and yields a constant:
\begin{equation}
	\int \d \Phi_2(P_1,P_2;q) = 2^{-3+2\e}\pi^{-1+\e}\dfrac{\Gamma(1-\e)}{\Gamma(2-2\e)}\left(q^2\right)^{-\e}.
\end{equation}
Considering now a process with $n$ final-state particles, the following factorisation holds:
\begin{eqnarray}\label{eq:PSfac}
	\d \Phi_n (p_1,.,p_{i_1},.,p_{i_{m}},.,p_n;q) = &&\d \Phi_{n-m+2} (p_1,.,P_{i_1},P_{i_2},\dots,p_{n-m+2};q)\nn\\
	&&\hspace{2.5cm}\cdot\,\d \Phi_{X_{m}}(p_{i_1},.,p_{i_m};P_{i_1}+P_{i_2}).
\end{eqnarray}
The antenna phase-space and the \textit{reduced} phase space in~\eqref{eq:PSfac} only cross-talk to each other via the mapped hard momenta $P_{i_1}$ and $P_{i_2}$, which are functions of the momenta living in the antenna phase space. Thanks to the factorisation above and the properties of the momentum mapping, one can perform the integration over the antenna phase space (and the potentially unresolved partons) independently from the fiducial selection criteria, which only affect resolved momenta in the reduced phase space. Different choices of momentum mappings do not affect the final result obtained after integration over the full fiducial phase space.

For massless particles, a general final-state momentum mapping algorithm addressing multiple singular emissions mapped onto two hard momenta, preserving momentum conservation and on-shellness conditions was presented in~\cite{Kosower:1997zr,Kosower:2002su}. Although first developed for tree-level gauge-theory amplitudes, this mapping, labelled antenna mapping, has been extensively applied for numerical integration in the context of the antenna subtraction method up to NNLO~\cite{Kosower:2003bh,Gehrmann-DeRidder:2005btv,Gehrmann-DeRidder:2007foh}.  In order to properly interpolate among different divergent configurations, a mapping has to properly reproduce the expected behaviour in soft and collinear limits. For example, considering a $m\to 2$ mapping, with $(m-2)$ unresolved momenta $p_3,\dots,p_m$, in the region $\soft{3,\dots,m}$ the mapping should give $P_1=p_1$ and $P_2=p_2$, while in the region $\coll{2,3,\dots,m}$ it should give $P_1=p_1$ and $P_2=p_{2}+p_{3}+\dots+ p_m$. Moreover, to successfully perform higher-order calculations with antenna functions, the employed momentum mapping is required to exactly transform into the mapping designed for lower unresolved multiplicity in less divergent final-state phase space. This reflects the fact that an antenna function is meant to simultaneously remove multiple infrared divergences across the phase space. For example, final-state $\{p^h_1,p_2,p_3,p^h_4\}\rightarrow\{P_{1},P_{2}\}$ and $\{p^h_1,p_2,p_3,p_4,p^h_5\}\rightarrow\{P_{1},P_{2}\}$ mappings should be equivalent to the $\{p^h_1,p_2,p^h_4\}\rightarrow\{P_{1},P_{2}\}$ mapping in  soft limits of $p_3\rightarrow 0$ and $p_3,p_4\rightarrow 0$. The mapping in~\cite{Kosower:1997zr,Kosower:2002su} satisfies all these properties. See~\cite{Kosower:2002su,Glover:2010kwr} for explicit proofs.

In general, however, the antenna mapping does not behave properly in all the possible unresolved configurations associated with an $m$-particle system. In particular, the mapping assumes a specific \textit{ordering} among the momenta and only reproduces the expected mapped momenta in those infrared limits which are allowed by the considered ordering. Soft emissions do not constitute an issue, because a soft momentum always disappears from the reconstructed hard momenta, while collinear clusters of particles which are not adjacent in the ordering assumed by the mapping are not properly treated. A detailed analysis is given below. The decomposition of the antenna phase space into sectors proposed in this paper precisely addresses this problem. We note that in the case of initial-final or initial-initial antenna functions for one or two initial-state hard radiators~\cite{Daleo:2006xa,Daleo:2009yj,Boughezal:2010mc,Gehrmann:2011wi} and in the case of identified final-state particles~\cite{Gehrmann:2022pzd,Bonino:2024adk} the preferred choice of mapping is an unordered dipole-like mapping~\cite{Catani:1996jh}, which better suits the requirement of keeping the direction of the initial-state or the identified particles unchanged after the mapping. Such a mapping does not exhibit the problem discussed above.

\subsubsection{Single-unresolved antenna mapping}
At NLO, only one singular emission is allowed. The $3\rightarrow 2$ antenna mapping $\{p^h_1,p_2,p^h_3\}\rightarrow\{P_1,P_2\}$, where $p_1^h$ and $p^h_3$ are the momenta of the hard ($h$) radiators, while $p_2$ is the momentum of the unresolved parton, is defined as~\cite{Kosower:1997zr,Kosower:2002su}:
\begin{eqnarray}
\label{eq:3to2mapping}
P^\mu_{1}&=&x\ p_1^\mu + r\ p_2^\mu + z\ p_3^\mu\nonumber\\
P^\mu_{2}&=&(1-x)\ p_1^\mu + (1-r)\ p_2^\mu + (1-z)\ p_3^\mu
\end{eqnarray}
where,
\begin{eqnarray}
x&=&\frac{1}{2(s_{12}+s_{13})}\big[(1+\rho)s_{123}-2r s_{23}\big]\\
z&=&\frac{1}{2(s_{23}+s_{13})}\big[(1-\rho)s_{123}-2 r s_{12}\big]\\
r&=&\frac{s_{23}}{s_{12}+s_{23}},\quad \rho^2=1+\frac{4 r(1-r)s_{12}s_{23}}{s_{123}s_{13}},
\end{eqnarray}
with $s_{ij}=2p_i\cdot p_j$. The two hard radiator momenta $p^h_1$ and $p^h_3$ are of equivalent role in the mapping. This mapping assumes the ordering $(1-2-3)$, hence it yields a suitable reconstruction of hard momenta in the reduced phase space in the following limits:
\begin{itemize}
	\item single-unresolved: $\soft{2}$, $\coll{1,2}$, $\coll{2,3}$,
\end{itemize}
which happen to be all the possible unresolved limits present in the $X_3^0$ antenna functions when partons $1$ and $3$ are considered hard. Therefore, the NLO antenna mapping works as intended in any infrared limit, for any NLO antenna function with well-defined hard radiators.

\subsubsection{Double-unresolved antenna mapping}\label{sec:du_map}
The $4\rightarrow 2$ antenna mapping $\{p^h_1,p_2,p_3,p^h_4\}\rightarrow\{P_{1},P_{2}\}$, for $p_1^h$ and $p^h_4$ hard radiators and $p_2$ and $p_3$ unresolved momenta, is given by~\cite{Kosower:2002su}:
\begin{eqnarray}
\label{eq:4to2mapping}
P^\mu_{1}&=&x\ p_1^\mu + r_1\ p_2^\mu + r_2\ p_3^\mu + z\ p_4^\mu\nonumber\\
P^\mu_{2}&=&(1-x)\ p_1^\mu + (1-r_1)\ p_2^\mu + (1-r_2)\ p_3^\mu + (1-z)\ p_4^\mu
\end{eqnarray}
where,
\begin{eqnarray}
r_1&=&\frac{s_{23}+s_{24}}{s_{12}+s_{23}+s_{24}}\nonumber\\
r_2&=&\frac{s_{34}}{s_{13}+s_{23}+s_{34}}\nonumber\\
x&=&\frac{1}{2(s_{12}+s_{13}+s_{14})}\bigg[(1+\rho)s_{1234}-r_1 (s_{23}+2s_{24})-r_2(s_{23}+s_{34}) \nonumber\\ &&+ (r_1-r_2)\frac{s_{12}s_{34}-s_{13}s_{24}}{s_{14}}\bigg]\nonumber\\
z&=&\frac{1}{2(s_{14}+s_{24}+s_{34})}\bigg[(1-\rho)s_{1234}-r_1 (s_{23}+2s_{12})-r_2(s_{23}+s_{13}) \nonumber\\ &&- (r_1-r_2)\frac{s_{12}s_{34}-s_{13}s_{24}}{s_{14}}\bigg]\nonumber
\end{eqnarray}
\begin{eqnarray}
\rho^2&=&1+\frac{(r_1-r_2)^2}{s_{14}^2s_{1234}^2}\lambda(s_{12}s_{34},s_{14}s_{23},s_{13}s_{24})\nonumber\\
&&+\frac{1}{s_{14}s_{1234}}\bigg[2\big(r_1(1-r_2)+r_2(1-r_1)\big)\big(s_{12}s_{34}+s_{13}s_{24}-s_{23}s_{14}\big)\nonumber\\
&&+4r_1(1-r_1)s_{12}s_{24}+4r_2(1-r_2)s_{13}s_{34}\bigg]\nonumber\\
\lambda(u,v,w)&=&u^2+v^2+w^2-2(uv+uw+vw).\nonumber
\end{eqnarray}
It assumes the ordering $(1-2-3-4)$ and therefore it yields a suitable reconstruction of hard momenta in the reduced phase space only in the following limits:
\begin{itemize}
	\item double-unresolved: 
	
	$\soft{2,3}$, 
	
	$\coll{1,2,3}$, $\coll{2,3,4}$, 
	
	$\soft{2}\otimes\coll{3,4}$, 
	$\soft{3}\otimes\coll{1,2}$, 
	$\soft{2}\otimes\coll{1,3}$, 
	$\soft{3}\otimes\coll{2,4}$, 
	
	$\coll{1,2}\otimes\coll{3,4}$;
	\item single-unresolved: $\soft{2}$, $\soft{3}$, $\coll{1,2}$, $\coll{2,3}$, $\coll{3,4}$;
\end{itemize}
while it fails for:
\begin{itemize}
	\item double-unresolved: $\coll{1,3}\otimes\coll{2,4}$;
	\item single-unresolved: $\coll{1,3}$, $\coll{2,4}$.
\end{itemize}
Namely, non-adjacent momenta are not allowed to become collinear, but in a collinear cluster with all the in-between momenta too. As mentioned, limits involving one or more soft emissions are always properly reconstructed, because soft particles drop out from the reconstructed hard momenta.  At NNLO, after the removal of one soft emission, one recovers an NLO-like three-particle configuration, where there are no ordering problems when the two hard radiators are fixed. As we will see, at N$^3$LO with three unresolved partons, the presence of a single soft emission is not enough to prevent the ordering problem.

Leading-colour four-particle antenna functions $X_4^0$ typically contain infrared limits associated to a specific colour ordering, hence an equally ordered mapping is suitable. However, beyond leading-colour, the $\wt{X}_4^0$ antenna functions, which are in general not associated to a specific colour ordering, can in principle contain unresolved configurations which violate the ordering required by the mapping. The prototypical example is the $\wt{A}_4^0(q,\tilde{g},\tilde{g},\bar{q})$ antenna function describing the emission of two abelian gluons (indicated with $\tilde{g}$) or photons between a quark-antiquark hard dipole.  In this case, the set of divergent limits in the antenna function cannot be fully covered by a single choice of ordering and the antenna mapping does not yield a correctly working subtraction term. As we illustrate in Section~\ref{sec:subantenna}, for previous applications of the antenna subtraction up to NNLO, the solution to this issue consisted in suitably splitting via partial fractioning the $\wt{X}_4^0$ antenna functions into \textit{sub-antennae} containing only selected divergences, in such a way that a specific ordered mapping can be used for each of them.

Generalised antenna functions with three hard radiators at NNLO induce a factorisation of the phase space which is different from the one in~\eqref{eq:antPS}-\eqref{eq:PSfac}~\cite{Fox:2024bfp}. The different factorisation calls for a different choice of momentum mapping to reabsorb the two unresolved emissions. In particular, iterated single-unresolved dipole mappings are used, for which no choice of a specific ordering is required. 

\subsubsection{Triple-unresolved antenna mapping}\label{subsubsec:3unresolvedmapping}
The $5\rightarrow2$ antenna mapping $\{p^h_1,p_2,p_3,p_4,p^h_5\}\rightarrow\{P_{1},P_{2}\}$ reads~\cite{Kosower:2002su}:
\begin{eqnarray}
\label{eq:5to2mapping}
P^\mu_{1}&=&x\ p_1^\mu + r_1\ p_2^\mu + r_2\ p_3^\mu + r_3\ p_4^\mu + z\ p_5^\mu\nonumber\\
P^\mu_{2}&=&(1-x)\ p_1^\mu + (1-r_1)\ p_2^\mu + (1-r_2)\ p_3^\mu + (1-r_3)\ p_4^\mu + (1-z)\ p_5^\mu
\end{eqnarray}
where,
\begingroup
\allowdisplaybreaks
\begin{eqnarray}
r_1&=&\frac{s_{23}+s_{24}+s_{25}}{s_{12}+s_{23}+s_{24}+s_{25}}\nonumber\\
r_2&=&\frac{s_{34}+s_{35}}{s_{13}+s_{23}+s_{34}+s_{35}}\nonumber\\
r_3&=&\frac{s_{45}}{s_{14}+s_{24}+s_{34}+s_{45}}\nonumber
\end{eqnarray}
\begin{eqnarray}
x&=&\frac{1}{2s_{15}(s_{12}+s_{13}+s_{14}+s_{15})}\bigg[r_3 s_{14}(s_{25}+s_{35})-r_3 s_{45} (s_{12}+s_{13}) + s_{23}s_{15}\nonumber\\
&&+s_{15}\big(s_{24} - r_{3}s_{24} + s_{12} + s_{25} + s_{34} + s_{13} + s_{35} + s_{14} + s_{45} \nonumber\\
&&+ \rho (s_{23} + s_{24} + s_{12} + s_{25} + s_{34} + s_{13} + s_{35} + s_{14} + s_{45}) - r_{3}(s_{34} + 2 s_{45})\big)\nonumber\\
&&+(1 + \rho)s_{15}^2 + r_2\big(-s_{35}(s_{12} + s_{14}) + s_{13}(s_{25} + s_{45}) - (s_{23} + s_{34} + 2 s_{35})s_{15}\big)\nonumber\\
&& - r_1\big(-s_{12}(s_{35} + s_{45}) + (s_{23} + s_{24})s_{15} + s_{25}(s_{13} + s_{14} + 2 s_{15})\big)\bigg]\nonumber\\
z&=&\frac{1}{2s_{15}(s_{25}+s_{35}+s_{45}+s_{15})}\bigg[-r_3 s_{14}(s_{25}+s_{35})+r_3 s_{45} (s_{12}+s_{13}) + s_{23}s_{15}\nonumber\\
&&+s_{15}\big(s_{24} - r_{3}s_{24} + s_{12} + s_{25} + s_{34} + s_{13} + s_{35} + s_{14} + s_{45} \nonumber\\
&&- \rho (s_{23} + s_{24} + s_{12} + s_{25} + s_{34} + s_{13} + s_{35} + s_{14} + s_{45}) - r_{3}(s_{34} + 2 s_{14})\big)\nonumber\\
&&+(1 - \rho)s_{15}^2 + r_1\big(s_{25}(s_{13} + s_{14}) - s_{12}(s_{35} + s_{45}) - (s_{23} + s_{24} + 2 s_{12})s_{15}\big)\nonumber\\
&& - r_2\big(-s_{35}(s_{12} + s_{14}) + (s_{23} + s_{34})s_{15} + s_{13}(s_{25} + s_{45} + 2 s_{15})\big)\bigg]\nonumber
\end{eqnarray}
\begin{eqnarray}
\rho^2&=&1 + \frac{2}{s_{15} s_{12345}} \bigg[-2 r_1^2 s_{12} s_{25} - 2 r_2^2 s_{13} s_{35} + r_1 \big\{s_{25} (s_{13} - 2 r_2 s_{13} + s_{14} - 2 r_3 s_{14}) \nonumber\\
&&\qquad+ s_{12} (2 s_{25} + s_{35} - 2 r_2 s_{35} + s_{45} - 2 r_3 s_{45}) + s_{15}(-s_{23} +  2 r_2 s_{23} - s_{24} + 2 r_3 s_{24}) \big\} \nonumber\\
&&\qquad+ r_2 \big\{s_{25} s_{13} + s_{12} s_{35} + s_{35} s_{14} - 2 r_3 s_{35} s_{14} + s_{13} (2 s_{35} + s_{45} - 2 r_3 s_{45}) + 2 r_3 s_{34} s_{15} \nonumber\\
&&\qquad- (s_{23} + s_{34}) s_{15}\big\} + r_3 \big\{(s_{12} + s_{13}) s_{45} + s_{14} (s_{25} + s_{35} - 2 (-1 + r_3) s_{45}) \nonumber\\
&&\qquad- (s_{24} + s_{34}) s_{15}\big\}\bigg] + \frac{1}{s_{15}^2 s_{12345}^2} \bigg[r_1^2 \big\{\big(s_{25} (s_{13} + s_{14}) - s_{12} (s_{35} + s_{45})\big)^2 \nonumber\\
&&\qquad- 2 s_{15}\big(-2 s_{12} s_{25} s_{34} + (s_{23} + s_{24}) s_{25} (s_{13} + s_{14}) + (s_{23} + s_{24}) s_{12} (s_{35} + s_{45})\big) \nonumber\\
&&\qquad+ (s_{23} + s_{24})^2 s_{15}^2\big\} + r_2^2 \big\{\big(s_{35} (s_{12} + s_{14}) - s_{13} (s_{25} + s_{45})\big)^2 \nonumber\\
&&\qquad - 2 s_{15}\big(s_{25} s_{34} s_{13} + s_{12} s_{34} s_{35} - 2 s_{24} s_{13} s_{35} + s_{34} s_{35} s_{14} + s_{23} s_{35} (s_{12} + s_{14})\nonumber\\
&&\qquad + s_{34} s_{13} s_{45} + s_{23} s_{13} (s_{25} + s_{45})\big) + (s_{23} + s_{34})^2 s_{15}^2\big\} \nonumber\\
&&\qquad+ r_3^2 \big\{\big((s_{25} + s_{35}) s_{14} - (s_{12} + s_{13}) s_{45}\big)^2 - 2 s_{15}\big((s_{24} + s_{34}) (s_{25} + s_{35}) s_{14} \nonumber\\
&&\qquad + (s_{24} + s_{34}) (s_{12} + s_{13}) s_{45} - 2 s_{23} s_{14} s_{45}\big)  + (s_{24} + s_{34})^2 s_{15}^2\big\} \nonumber\\
&&\qquad - 2 r_1 \big\{\big(r_3 (s_{25} + s_{35}) s_{14} - r_2 s_{35} (s_{12} + s_{14}) -  r_3 (s_{12} + s_{13}) s_{45} + r_2 s_{13} (s_{25} + s_{45})\big)\nonumber\\
&&\qquad \times \big(s_{25} (s_{13} + s_{14}) - s_{12} (s_{35} + s_{45})\big) + s_{15}\big(2 r_2 s_{12} s_{25} s_{34} + 2 r_3 s_{12} s_{25} s_{34} \nonumber\\
&&\qquad+ r_3 s_{25} (-s_{24} + s_{34}) s_{13} - r_2 s_{25} (2 s_{23} + s_{24} + s_{34}) s_{13} + r_2 s_{25} (-s_{23} + s_{34}) s_{14} \nonumber\\
&&\qquad- r_3 s_{25} (s_{23} + 2 s_{24} + s_{34}) s_{14}\nonumber\\ 
&&\qquad- r_2 s_{35} \big(s_{12} s_{34} + s_{24} (s_{12} - 2 s_{13} - s_{14}) + s_{23} (2 s_{12} + s_{14})\big)\nonumber\\
&&\qquad + r_3 s_{35} \big(s_{12} s_{34} + s_{23} s_{14} - s_{24} (s_{12} + 2 s_{13} + s_{14})\big) \nonumber\\
&&\qquad - r_3 \big(s_{12} s_{34} + s_{24} (2 s_{12} + s_{13}) + 
              s_{23} (s_{12} - s_{13} - 2 s_{14})\big) s_{45}\nonumber\\
&&\qquad+ r_2 (s_{12} s_{34} + s_{24} s_{13} - 
              s_{23} (s_{12} + s_{13} + 2 s_{14})) s_{45}\big)\nonumber\\
&&\qquad+ s_{15}^2\big((s_{23} + 
              s_{24}) (r_2 s_{23} + r_3 s_{24}) + (r_2 - r_3) (s_{23} - 
              s_{24}) s_{34}\big) \big\} \nonumber\\
&&\qquad+ 2 r_2 r_3 \big\{s_{25}^2 s_{13} s_{14} + s_{12}^2 s_{35} s_{45} - s_{12} (s_{35} - s_{45}) (s_{35} s_{14} - s_{13} s_{45}) \nonumber\\
&&\qquad - (s_{35} s_{14} - s_{13} s_{45})^2 + s_{15}\big(s_{24} s_{35} s_{14} + (s_{23} + 2 s_{34}) s_{35} s_{14} + 2 s_{34} s_{13} s_{45}\nonumber\\
&&\qquad + s_{23} (s_{13} - 2 s_{14}) s_{45} + s_{24} s_{13} (-2 s_{35} + s_{45})\big) \nonumber\\
&&\qquad + s_{12} \big(-s_{24} s_{35} - s_{23} s_{45} + s_{34} (s_{35} + s_{45})\big) s_{15}\nonumber\\
&&\qquad + \big(s_{23} (s_{24} - s_{34}) - 
           s_{34} (s_{24} + s_{34})\big) s_{15}^2 + 
        s_{25} \big(-s_{35} s_{14} (s_{12} - s_{13} + s_{14})\nonumber\\
&&\qquad - s_{13} (s_{12} + s_{13} - s_{14}) s_{45} + (-s_{24} s_{13} - s_{23} s_{14} + 
              s_{34} (2 s_{12} + s_{13} + s_{14})) s_{15}\big)\big\}\bigg].\nonumber
\end{eqnarray}
\endgroup

This mapping assumes the ordering $(1-2-3-4-5)$ and therefore it yields a suitable reconstruction of hard momenta in the reduced phase space only in the following unresolved limits:
\begin{itemize}
	\item triple-unresolved: 
	
	$\soft{2,3,4}$,
	 
	$\soft{2,3}\otimes\coll{4,5}$, $\soft{2,3}\otimes\coll{1,4}$, 
	$\soft{3,4}\otimes\coll{1,2}$, $\soft{3,4}\otimes\coll{2,5}$, 
	$\soft{2,4}\otimes\coll{1,3}$, $\soft{2,4}\otimes\coll{3,5}$, 
	
	$\soft{2}\otimes\coll{3,4,5}$, $\soft{2}\otimes\coll{1,3,4}$, 
	$\soft{4}\otimes\coll{1,2,3}$, $\soft{4}\otimes\coll{2,3,5}$, 
	$\soft{3}\otimes\coll{1,2,4}$, $\soft{3}\otimes\coll{2,4,5}$,
	
	$\soft{2}\otimes\coll{1,3}\otimes\coll{4,5}$, 
	$\soft{4}\otimes\coll{1,2}\otimes\coll{3,5}$, 
	$\soft{3}\otimes\coll{1,2}\otimes\coll{4,5}$, 
	
	$\coll{1,2,3,4}$, $\coll{2,3,4,5}$;
	\item double-unresolved: 
	
	$\soft{2,3}$, $\soft{2,4}$, $\soft{3,4}$,
	 
	$\soft{2}\otimes\coll{3,4}$, 
	$\soft{2}\otimes\coll{4,5}$, 
	$\soft{2}\otimes\coll{1,3}$, 
	$\soft{3}\otimes\coll{1,2}$, 
	$\soft{3}\otimes\coll{4,5}$,\\ 
	$\soft{3}\otimes\coll{2,4}$, 
	$\soft{4}\otimes\coll{1,2}$, 
	$\soft{4}\otimes\coll{2,3}$, 
	$\soft{4}\otimes\coll{3,5}$, 
	
	$\coll{1,2}\otimes\coll{3,4}$, 
	$\coll{2,3}\otimes\coll{4,5}$, 
	$\coll{1,2}\otimes\coll{4,5}$,
	
	$\coll{1,2,3}$, $\coll{2,3,4}$, $\coll{3,4,5}$;
	\item single-unresolved: 
	$\soft{2}$, $\soft{3}$, $\soft{4}$, 	
	$\coll{1,2}$, $\coll{2,3}$, $\coll{3,4}$, $\coll{4,5}$,
\end{itemize}
while it fails for:
\begin{itemize}
	\item triple-unresolved: 
	
	$\soft{2}\otimes\coll{1,4}\otimes\coll{3,5}$, 
	$\soft{3}\otimes\coll{1,4}\otimes\coll{2,5}$, 
	$\soft{4}\otimes\coll{1,3}\otimes\coll{2,5}$, 
	
	$\coll{1,2,4}\otimes\coll{3,5}$, 
	$\coll{1,3,4}\otimes\coll{2,5}$, 
	$\coll{1,3}\otimes\coll{2,4,5}$,  
	$\coll{1,4}\otimes\coll{2,3,5}$;
	\item double-unresolved: 
	
	$\coll{1,2,4}$, 
	$\coll{1,3,4}$, 
	$\coll{2,3,5}$,  
	$\coll{2,4,5}$,
	 
	$\coll{1,2}\otimes\coll{3,5}$, 
	$\coll{1,3}\otimes\coll{2,5}$, 
	$\coll{1,3}\otimes\coll{4,5}$, 
	$\coll{1,3}\otimes\coll{2,4}$,\\ 
	$\coll{1,4}\otimes\coll{2,5}$, 
	$\coll{1,4}\otimes\coll{3,5}$, 
	$\coll{1,4}\otimes\coll{2,3}$, 
	$\coll{2,4}\otimes\coll{3,5}$,\\ 
	$\coll{3,4}\otimes\coll{2,5}$,
	
	$\soft{2}\otimes\coll{3,5}$, 
	$\soft{2}\otimes\coll{1,4}$, 
	$\soft{3}\otimes\coll{1,4}$, 
	$\soft{3}\otimes\coll{2,5}$, 
	$\soft{4}\otimes\coll{1,3}$, \\
	$\soft{4}\otimes\coll{2,5}$;
	\item single-unresolved: $\coll{1,3}$, $\coll{1,4}$, $\coll{2,5}$, $\coll{3,5}$, $\coll{2,4}$.
\end{itemize}
As the number of unresolved emissions increases, the antenna mapping fails to reconstruct the hard momenta in a handful of configurations. Clearly, the sub-antenna strategy can be extended at N$^3$LO, by partial-fractioning the problematic antenna functions to isolate divergences which can be addressed with a specific ordered mapping. However, as mentioned above, this is neither simple nor elegant and potentially introduces spurious large cancellations in the numerical implementation.


\subsection{The sub-antenna functions}\label{sec:subantenna}

At NNLO, the solution to the problem of assigning an ordered momentum mapping to a generic antenna functions has been addressed by the definition of sub-antennae. In this section we briefly summarise their definition to illustrate the cases which require them.   Sub-antennae are obtained by partial fractioning to separate singular denominators. The obtained pieces contain only some of the divergences  present in the original antenna function, thus an ordered mapping can be assigned to them. Sub-antennae have been defined for the sake of  isolating a specific choice of hard radiators (for traditional antenna functions) or to split singularities arising from multiple unordered emissions (for traditional and designer antenna functions).

Repeated partial fractioning leads to larger and more complicated expressions for the sub-antennae. The presence of unphysical propagators in the sub-antennae makes their analytical integration harder, typically requiring new master integrals with respect to those needed for the original antenna functions. However, as observed in~\cite{Gehrmann-DeRidder:2005btv}, one can avoid the analytical integration of the individual sub-antennae relying on the fact that they induce the same phase-space factorisation and can therefore always be recombined together after integration in the reduced phase-space. A related issue consists in the fact that \textit{all} the divergent terms in a given antenna function get re-distributed between its sub-antennae, even those, like soft-divergent terms, which would have not required partial-fractioning. This may cause large cancellations in the numerical implementation of the subtraction terms. We observe that at NNLO, where sub-antennae are used for the double-real subtraction terms, for calculations with a well-resolved underlying  Born configuration, this never constitutes an issue for the numerics. However, at N$^3$LO, both in the case of triple-real emissions, where the number of singular denominators proliferates, or double-real-virtual antenna functions, where non-rational and less numerically stable functions appear, construction and application of sub-antennae become impractical. 

In the rest of this section, we summarise the partonic configurations which required the definition of sub-antennae, both for the traditional antenna functions and the designer ones. The latter will be accompanied by a $des.$ superscript to distinguish them.

At NLO, where a single emission occurs, the only ambiguity is given by the choice of hard radiators in antenna functions with multiple gluons. This affects the traditional matrix element based quark-gluon antenna function $D_3^0$ and gluon-gluon antenna function $F_3^0$, respectively describing the emission of a gluon from a quark-gluon and gluon-gluon dipole~\cite{Gehrmann-DeRidder:2005alt}.  The full antenna function $D_3^0(q,g_1,g_2)$ can be split into two sub-antennae:
\begin{equation}
	D_3^0(q,g_1,g_2)=d_3^0(q,g_1,g_2)+d_3^0(q,g_2,g_1),
\end{equation} 
where gluon $g_1$ ($g_2$) is identified as the hard radiator and $g_2$ ($g_1$) as the radiated parton which can become soft. One can choose how to distribute the collinear limit $g_1\parallel g_2$ between the sub-antennae~\cite{Gehrmann-DeRidder:2005btv}. Analogously, the $F_3^0$ antenna function can be decomposed according to: 
\begin{equation}
	F_3^0(g_1,g_2,g_3)=f_3^0(g_1,g_2,g_3)+f_3^0(g_2,g_3,g_1)+f_3^0(g_3,g_1,g_2),
\end{equation}
where each sub-antenna $f_3^0(g_i,g_j,g_k)$ specifically targets the soft limit of gluon $g_j$ and has $g_i$ and $g_k$ as hard radiators. For the sub-antennae $d_3^0$ and $f_3^0$ one can choose how to distribute the collinear limits, provided a coherent choice is made between the two types of antenna functions. Specific expressions for the sub-antennae are given in~\cite{Gehrmann-DeRidder:2005btv}. The antenna functions $d_3^0$ and $f_3^0$ can also be constructed and individually integrated in the designer antenna approach~\cite{Braun-White:2023sgd}, where the choice of hard radiators is fixed from the beginning.

Problems related to multiple unordered emissions first appear at NNLO. We start by considering the quark-antiquark antenna function $\wt{A}_4^0(q,g_1,g_2,\qb)$. It describes the emission of two abelian gluons from a hard quark-antiquark pair~\cite{Gehrmann-DeRidder:2004ttg,Gehrmann-DeRidder:2005alt}. The analogous idealised antenna function $\wt{A}_4^{0,des.}(q^h,g_1,g_2,\qb^h)$ has the same infrared structure~\cite{Braun-White:2023sgd}, hence anything we explain in the following holds for it too. No notion of colour ordering is present: all the quark-gluon collinear limits are singular, both the ones naturally present in the ordering above, $\coll{q,g_1}$, $\coll{g_2,\bar{q}}$ and the others $\coll{q,g_2}$, $\coll{g_1,\bar{q}}$. This makes it impossible to associate a single ordered mapping to the $\wt{A}_4^0(q,g_1,g_2,\qb)$ antenna function. To solve this, it is then be split into two sub-antennae~\cite{Gehrmann-DeRidder:2007foh}:
\begin{equation}
	\wt{A}_4^0(q,g_1,g_2,\qb)=\wt{A}_{4,a}^0(q,g_1,g_2,\qb)+\wt{A}_{4,a}^0(q,g_2,g_1,\qb),
\end{equation}
which only retain specific singular limits and therefore act as colour-ordered antenna functions. In particular, $\wt{A}_{4,a}^0(q,g_1,g_2,\qb)$ is obtained from the full antenna functions separating the terms coming with negative powers of $s_{qg_1}$ and $s_{g_2\qb}$ from the ones with negative powers of $s_{qg_2}$ and $s_{g_1\qb}$, and retaining only the divergent behaviour in the $\coll{q,g_1}$ and $\coll{g_2,\bar{q}}$ limits.  Therefore, all the infrared limits of $\wt{A}_{4,a}^0(q,g_1,g_2,\qb)$ are well addressed by a momentum mapping with ordering $(q-g_1-g_2-\qb)$. The very same discussion holds for the counterpart antenna function obtained with the designer algorithm $\wt{A}_{4}^{0,des.}$.

We continue with the quark-gluon $D_4^0(q,g_1,g_2,g_3)$ antenna function, describing the emission of two gluons from a hard quark-gluon pair~\cite{GehrmannDeRidder:2005hi,Gehrmann-DeRidder:2005btv}. Here, both the problem of ambiguous hard radiators and unordered emissions occurs. It can be decomposed according to~\cite{Gehrmann-DeRidder:2007foh}: 
\begin{eqnarray}
	D_4^0(q,g_1,g_2,g_3) &=& D_{4,a}^0(q^h,g_1,g_2,g^h_3) + D_{4,a}^0(q^h,g_3,g_2,g^h_1) \nonumber \\
	&+& D_{4,c}^0(q^h,g_1,g^h_2,g_3) + D_{4,c}^0(q^h,g_3,g^h_2,g_1),
\end{eqnarray}
where the superscript $h$ indicates the choice of hard radiators. $D^0_{4,a}$ is a genuine colour-ordered contributions, while the two $D^0_{4,c}$ can be obtained from the leftover by a partial-fractioning procedure analogous to the one applied to define the $\wt{A}^0_{4,a}$ sub-antenna, namely distinguishing which gluon ($g_1$ or $g_3$) is allowed to be collinear to the hard quark (gluon). The separation into $D^0_{4,a}$ and $D^0_{4,c}$ is more clear from the perspective of the idealised antenna function algorithm~\cite{Braun-White:2023sgd}, according to which $D^0_{4,a}$ and $D^0_{4,c}$ are associated to different sub-structures present in the double-unresolved factors. As far as infrared divergences are concerned, the following relation hold:
\begin{eqnarray}
	D_4^{0,des.}(q^h,g_1,g_3,g_2^h)&\sim& D_4^{0,a}(q,g_1,g_2,g_3),\\
	\wt{D}_4^{0,des.}(q^h,g_1,g_3,g_2^h)&\sim& D_4^{0,c}(q,g_1,g_2,g_3) +  D_4^{0,c}(q,g_3,g_2,g_1).
\end{eqnarray}
From the second relation above, it is clear that even the idealised antenna function $\wt{D}_4^{0,des.}$ would need to be split into two sub-antennae in order to correctly assign a single choice of ordered mapping to it. 

We consider then the gluon-gluon antenna function $F_4^0(g_1,g_2,g_3,g_4)$, describing the emission of two gluons between a pair of hard gluon radiators. The full antenna function is symmetric over the exchange of any pair of gluons and contains all possible unresolved limits and choices of hard radiators. It can be decomposed according to~\cite{Glover:2010kwr}:
\begin{eqnarray}
	F_4^0(g_1,g_2,g_3,g_4) &=& F_{4,a}^0(g_1^h,g_2,g_3,g_4^h) + F_{4,b}^0(g_1^h,g_2,g_3^h,g_4) \nonumber \\
	&+& F_{4,a}^0(g_1^h,g_4,g_3,g_2^h) + F_{4,b}^0(g_1^h,g_4,g_3^h,g_2) \nonumber \\
	&+& F_{4,b}^0(g_2^h,g_3,g_4^h,g_1) + F_{4,a}^0(g_2^h,g_1,g_4,g_3^h) \nonumber \\
	&+& F_{4,b}^0(g_4^h,g_3,g_2^h,g_1) + F_{4,a}^0(g_4^h,g_1,g_2,g_3^h).
\end{eqnarray}
As for the $D_{4,a}^0$ case above, the $F_{4,a}^0$ sub-antennae is colour-ordered, while the $F_{4,b}^0$ terms are obtained by partial-fractioning the unordered leftover. These two contributions are naturally separated in the idealised antenna approach~\cite{Braun-White:2023sgd}. In terms of infrared limits, the following relations hold:
\begin{eqnarray}
	F_4^{0,des.}(g_1^h,g_2,g_3,g_4^h)&\sim &F_4^{0,a}(g_1^h,g_2,g_3,g_4^h),\\
	\wt{F}_4^{0,des.}(g_1^h,g_2,g_3,g_4^h)&\sim& F_4^{0,b}(g_1^h,g_2,g_3,g_4^h)+F_4^{0,b}(g_1^h,g_3,g_2,g_4^h).
\end{eqnarray}
Once again, the idealised antenna function $\wt{F}_4^{0,des.}$ would need to be split into two contributions via partial-fractioning to identify a specific ordered momentum mapping.

Finally, we mention that the quark-gluon antenna function $E_4^0(q,q',\bar{q}',g)$ and the gluon-gluon antenna function $G_4^0(g_1,q,\bar{q},g_2)$ also have to be split in sub-antennae, purely for the choice of hard radiators (between one of the secondary quarks and a gluon). However, such ambiguity is fully solved in the context of designer antenna functions~\cite{Braun-White:2023sgd}, with no further issue related to the presence of unordered emissions.

\vspace{0.5cm}

The sub-antennae strategy provides a practical way to obtain a working implementation of the antenna subtraction method and has been fully successful at NNLO, but does come with drawbacks which can become relevant at higher orders. In the following we present an alternative solution which scales much better with the number of unresolved emissions, does not require dedicated work (partial fractioning) for individual antenna functions and is less prone to numerical instabilities. 

\section{Phase space sectors}\label{sec:sectors}

In this section we present an algorithmic decomposition of the phase space aimed at isolating regions where only the infrared limits associated to a specific colour ordering exist. Such implementation bypasses the need of sub-antennae: full antenna functions can be associated to a given choice of ordered mapping according to the phase-space sector each event belongs to. The fact that the factorisation of the phase space and the leftover integration over the reduced phase space are unaffected by the choice of mapping guarantee that one recovers the correct result after integration over the whole phase space. 

\subsection{Double-unresolved sectors}\label{sec:2unsectors}

Problems with the assignment of a unique ordered mapping to antenna functions are first encountered at NNLO and in particular in double-real subtraction terms. As discussed in Section~\ref{sec:subantenna}, several $X_4^0$ antenna functions need to be split first of all to identify unambiguously two hard radiators. Thanks to the designer antenna algorithm~\cite{Braun-White:2023sgd}, we can however neglect this problem and assume that each antenna function has two well-defined hard radiators. This leaves us with the problem of singularities due to the emission of two unordered abelian gluons. Antenna functions with such singularities are $\wt{A}_4^0$, $\wt{A}_4^{0,des.}$, $D_{4,c}^{0}$, $\wt{D}_{4}^{0,des.}$, $F_{4,b}^{0}$ and $\wt{F}_{4}^{0,des.}$.  

We indicate these antenna functions with $\wt{X}_4^0(p_1^h,\tilde{g}_2,\tilde{g}_3,p^h_4)$, where $p^h_i$ are generic hard radiators (quark or gluons) and $\tilde{g}_i$ are the abelian gluons. The colour connection is $(1-2-4)\oplus(1-3-4)$. Since both abelian gluons can become collinear with either of the hard radiators, no unique ordered mapping can be associated to such configuration. However, one can split up the phase space into two regions, one (region $a$) where only the $\coll{1,2}$ and $\coll{3,4}$ infrared limits exist and one (region $b$) where only the $\coll{1,3}$ and $\coll{2,4}$ limits exist. It is clear that in region $a$ the $\wt{X}_4^0(p_1^h,\tilde{g}_2,\tilde{g}_3,p^h_4)$ can be treated as a fully ordered antenna function with colour connections $(1-2-3-4)$. The opposite holds for region $b$ and the ordering $(1-3-2-4)$. 

To define the two regions and choose the appropriate momentum mapping, we proceed in the following way:
\begin{enumerate}
\item For a given set of momentum $\{p^h_1,p_2,p_3,p^h_4\}$ entering a $\wt{X}_4^0(p_1^h,\tilde{g}_2,\tilde{g}_3,p^h_4)$ antenna function, we evaluate the products of Mandelstam invariants $s_{12}s_{34}$ and $s_{13}s_{24}$.

\item The full phase space is then split into two sectors, where a specific choice of ordered antenna mapping is used:
\begin{enumerate}
\item \label{sector:RR1} $s_{12}s_{34} \le s_{13}s_{24}$: $\{4\rightarrow 2\}$ mapping with ordering $\{p^h_1,p_2,p_3,p^h_4\}$. 

\item \label{sector:RR2} $s_{12}s_{34} > s_{13}s_{24}$: $\{4\rightarrow 2\}$ mapping with ordering $\{p^h_1,p_3,p_2,p^h_4\}$. 
\end{enumerate}

\item The evaluation of the $\wt{X}_4^0$ antenna functions is done with the unmapped momenta as usual. The reduced matrix element (or secondary antenna functions for calculations beyond NNLO) is evaluated on the mapped momenta corresponding to the appropriate mappings.
\end{enumerate}

The algorithm relies on the fact that, for an NNLO calculation, the conditions on the Mandelstam invariants only allow some of them to vanish in each region. We postpone to Section~\ref{sec:validation} analytical and numerical proofs of the correctness of the procedure described above. We notice that the algorithm is particularly simple and completely removes the necessity of sub-antennae for unordered emissions. Any infrared limit present in the $\wt{X}_4^0$ antenna function is always correctly reconstructed thanks to the discrimination between the two regions. For each phase-space point, the antenna function is evaluated in its entirety. This means that the soft terms, or any other divergent term that would not have required the separations into sub-antennae, does not undergo large cancellations.

The phase-space sectors defined above for $\wt{X}_4^0$, straightforwardly apply for double-unresolved one-loop antenna functions $\wt{X}_4^1$ too, which are needed for N$^3$LO calculations. This is another direct advantage of the strategy proposed here, namely no additional work is required at higher orders. In~\cite{Chen:2025kez}, the sectors above have been employed to assign a unique momentum mapping in the numerical implementation of the $\wt{A}_4^1$ antenna function, describing the emission of two potentially unresolved abelian gluons between a hard quark-antiquark pair at one-loop.

\subsection{Triple-unresolved sectors}\label{sec:3unsector}

With three unresolved emissions, needed for N$^3$LO calculations, there are multiple scenarios which require dedicated phase-space sectors to identify a unique ordered momentum mapping. 

First of all, one can have a fully unordered set of emissions, where any of the collinear limits involving an unresolved parton and either hard radiator is infrared divergent. A second scenario, which appears only with three or more unresolved partons, consists in a mixture of ordered and unordered emissions. In particular, at N$^3$LO an abelian gluon can be emitted in association with two ordered gluons, which can only become collinear to the adjacent hard radiators (and to each other). The configuration with two unordered gluons is equivalent to the first case, since there is no ordering for a single non-abelian gluon. Finally, for some partonic configurations at N$^3$LO, it may not be straightforward to disentangle in the definition of unresolved factors contributions due to the emission of a gluon between different dipoles. In this case thus, the unresolved factors can be written as a sum of terms with different colour connections, hence orderings. Clearly, by carefully separating singular denominators one can in principle extract unresolved factors with a unique pattern of gluonic emission, however we will show that by introducing dedicated phase-space sectors it is possible to entirely avoid this. 

We will now discuss the possibilities above, referring to explicit examples in the context of the recent calculation of the N$^3$LO correction to jet production at electron-positron colliders with antenna subtraction~\cite{Chen:2025kez}. Beyond N$^3$LO more possibilities need to be considered (\textit{e. g.} two unordered and two ordered emissions), but we will not address them here for the lack of practical necessity. Nevertheless, we argue that addressing the mapping problem with the implementation discussed in this paper is quite straightforward at any perturbative order.

\subsubsection{Three unordered emissions}

We start by considering the emission of three abelian gluons from a hard dipole. The prototypical example is again given by a hard quark-antiquark pair emitting three abelian gluons, described by the $\vardbtilde{A}_5^0(q^h_1,\tilde{g}_2,\tilde{g}_3,\tilde{g}_4,\bar{q}^h_5)$ triple-real antenna function. If one or both of the hard radiators are replaced by a gluon, one would obtain  the $\vardbtilde{D}_5^0$ and the  $\vardbtilde{F}_5^0$ antenna functions, analogously to the NNLO case. We generically indicate this class of antenna functions with $\vardbtilde{X}_5^0(p^h_1,\tilde{g}_2,\tilde{g}_3,\tilde{g}_4,p^h_5)$. The corresponding colour connection is $(1-2-5)\oplus(1-3-5)\oplus(1-4-5)$, with no unique ordering to be identified for the mapping. 

At NNLO, with two unresolved emissions, only two orderings exist and so two phase-space regions are enough to distinguish them. Now with three unordered emissions, $3!=6$ orderings are possible. We design an algorithmic procedure which splits the phase space into $12$ different sectors containing a subset of infrared configurations which can be addressed by one of the $6$ ordered mapping. The procedure reads as follow:
\begin{enumerate}
\item For a given set of momentum $\{p^h_1,p_2,p_3,p_4,p^h_5\}$ entering a $\vardbtilde{X}_5^0(p^h_1,\tilde{g}_2,\tilde{g}_3,\tilde{g}_4,p^h_5)$ antenna function, we first evaluate $s_{hi}=min\{s_{12},s_{13},s_{14},s_{25},s_{35},s_{45}\}$ (with $h\in\{1,5\}$ and $i\in\{2,3,4\}$) and $p_i$ is removed from the set of momenta, obtaining $\{p^h_1,p_j,p_k,p^h_5\}$. We then evaluate the products of Mandelstam invariants $s_{1j}s_{5k}$ and $s_{1k}s_{5j}$ (similarly to the double-unresolved case).

\item The full phase space is split into 12 sectors, where a specific choice of the 6 possible antenna mappings is used:
\begin{enumerate}
\item $s_{hi}=s_{12}$ and $s_{13}s_{45} \le s_{14}s_{35}$: $\{5\rightarrow 2\}$ mapping with ordering $\{p^h_1,p_2,p_3,p_4,p^h_5\}$;
\item $s_{hi}=s_{45}$ and $s_{12}s_{35} \le s_{13}s_{25}$: $\{5\rightarrow 2\}$ mapping with ordering $\{p^h_1,p_2,p_3,p_4,p^h_5\}$;
\item $s_{hi}=s_{14}$ and $s_{13}s_{25} \le s_{12}s_{35}$: $\{5\rightarrow 2\}$ mapping with ordering $\{p^h_1,p_4,p_3,p_2,p^h_5\}$;
\item $s_{hi}=s_{25}$ and $s_{14}s_{35} \le s_{13}s_{45}$: $\{5\rightarrow 2\}$ mapping with ordering $\{p^h_1,p_4,p_3,p_2,p^h_5\}$;
\item $s_{hi}=s_{13}$ and $s_{12}s_{45} \le s_{14}s_{25}$: $\{5\rightarrow 2\}$ mapping with ordering $\{p^h_1,p_3,p_2,p_4,p^h_5\}$;
\item $s_{hi}=s_{45}$ and $s_{12}s_{35} > s_{13}s_{25}$: $\{5\rightarrow 2\}$ mapping with ordering $\{p^h_1,p_3,p_2,p_4,p^h_5\}$;
\item $s_{hi}=s_{14}$ and $s_{13}s_{25} > s_{12}s_{35}$: $\{5\rightarrow 2\}$ mapping with ordering $\{p^h_1,p_4,p_2,p_3,p^h_5\}$;
\item $s_{hi}=s_{35}$ and $s_{14}s_{25} \le s_{12}s_{45}$: $\{5\rightarrow 2\}$ mapping with ordering $\{p^h_1,p_4,p_2,p_3,p^h_5\}$;
\item $s_{hi}=s_{12}$ and $s_{13}s_{45} > s_{14}s_{35}$: $\{5\rightarrow 2\}$ mapping with ordering $\{p^h_1,p_2,p_4,p_3,p^h_5\}$;
\item $s_{hi}=s_{35}$ and $s_{14}s_{25} > s_{12}s_{45}$: $\{5\rightarrow 2\}$ mapping with ordering $\{p^h_1,p_2,p_4,p_3,p^h_5\}$;
\item $s_{hi}=s_{13}$ and $s_{12}s_{45} > s_{14}s_{25}$: $\{5\rightarrow 2\}$ mapping with ordering $\{p^h_1,p_3,p_4,p_2,p^h_5\}$;
\item $s_{hi}=s_{25}$ and $s_{14}s_{35} > s_{13}s_{45}$: $\{5\rightarrow 2\}$ mapping with ordering $\{p^h_1,p_3,p_4,p_2,p^h_5\}$.
\end{enumerate}

\item The evaluation of the $\vardbtilde{X}_5^0$ antenna functions is done with the unmapped momenta as usual. The reduced matrix element is evaluated on the mapped momenta corresponding to the appropriate mapping.
\end{enumerate}
The underlying logic is the following: by individuating the minimum invariant $s_{hi}$, we can first position one of the potentially unresolved momenta close to a hard radiator; subsequently the ordering of the two remaining emissions is chosen by applying the same procedure used in the double-unresolved case.  The sectors defined above can be split into $6$ pairs by the choice of $s_{hi}$. The two sectors in each pair cover complementary slices of phase space (\textit{e.g.} sector (a) and (i)). Hence, the 12 sectors clearly do not overlap and cover the full phase space. 

\subsubsection{One unordered emission and two ordered ones}

We consider the next case where two of the three emissions are ordered, while the third is not. The prototypical example is the $\wt{A}_5^0(q^h_1,\tilde{g}_2,g_3,g_4,\bar{q}^h_5)$ triple-real antenna function, which describe the emission of two ordered gluons ($g_3$ and $g_4$) and a abelian one ($\tilde{g}_2$) from a hard quark-antiquark pair. Again, if one or both hard radiators are gluons, one would obtains the $\wt{D}_5^0$ and the  $\wt{F}_5^0$ antenna functions. We generically indicate this class of antenna functions with $\wt{X}_5^0(p^h_1,\tilde{g}_2,g_3,g_4,p^h_5)$. The corresponding colour connection is and $(1-2-5)\oplus(1-3-4-5)$.

Since the order of two of the emissions between the hard radiators is fixed, only three orderings are possible, corresponding to three different positions for the abelian gluon. We design an algorithmic procedure which splits the phase space into $12$ different sectors containing a subset of infrared configurations which can be addressed by one of the $3$ ordered mapping. The procedure reads as follow:
\begin{enumerate}
\item For a given set of momenta $\{p^h_1,p_2,p_3,p_4,p^h_5\}$ in $\wt{X}_5^0(p^h_1,\tilde{g}_2,g_3,g_4,p^h_5)$, where $p_2$ is the momentum of abelian gluon, we first evaluate $s_{hi}=min\{s_{12},s_{13},s_{14},s_{25},s_{35},s_{45}\}$ (with $h\in\{1,5\}$ and $i\in\{2,3,4\}$) and we compare it with $s_{34}$. If $s_{34} < s_{hi}$, we discard the ordering of $\{p^h_1,p_3,p_2,p_4,p^h_5\}$ and further evaluate the products of Mandelstam invariants $s_{1(3+4)}s_{25}$ and $s_{12}s_{(3+4)5}$, where $(3+4)$ indicates a cluster with momentum $p_3+p_4$. If $s_{34} \ge s_{hi}$ and $i=3,4$, we resolve the residual ambiguity by evaluating $s_{1j}s_{25}$ and $s_{12}s_{j5}$, with $j\ne 2,i$.

\item The full phase space is split into 12 sectors, where a specific choice of the three possible antenna mappings is used:
\begin{enumerate}

\item $s_{hi}\ge s_{34}$,
\begin{enumerate}
  \item $s_{1(3+4)}s_{25} \le s_{12}s_{(3+4)5}$: $\{5\rightarrow 2\}$ mapping with ordering $\{p^h_1,p_3,p_4,p_2,p^h_5\}$;
  \item $s_{1(3+4)}s_{25} > s_{12}s_{(3+4)5}$: $\{5\rightarrow 2\}$ mapping with ordering $\{p^h_1,p_2,p_3,p_4,p^h_5\}$;
\end{enumerate}

\item $s_{hi} < s_{34}$ and $s_{hi}=s_{12}$: $\{5\rightarrow 2\}$ mapping with ordering $\{p^h_1,p_2,p_3,p_4,p^h_5\}$.;

\item $s_{hi} < s_{34}$ and $s_{hi}=s_{25}$: $\{5\rightarrow 2\}$ mapping with ordering $\{p^h_1,p_3,p_4,p_2,p^h_5\}$;

\item $s_{hi} < s_{34}$ and $s_{hi}=s_{13}$,
\begin{enumerate}
\item $s_{14}s_{25} > s_{12}s_{45}$: $\{5\rightarrow 2\}$ mapping with ordering $\{p^h_1,p_3,p_2,p_4,p^h_5\}$; 
\item $s_{14}s_{25} \le s_{12}s_{45}$: $\{5\rightarrow 2\}$ mapping with ordering $\{p^h_1,p_3,p_4,p_2,p^h_5\}$; 
\end{enumerate}

\item $s_{hi} < s_{34}$ and  $s_{hi}=s_{14}$,
\begin{enumerate}
\item $s_{13}s_{25} > s_{12}s_{35}$: $\{5\rightarrow 2\}$ mapping with ordering $\{p^h_1,p_2,p_3,p_4,p^h_5\}$;
\item $s_{13}s_{25} \le s_{12}s_{35}$: $\{5\rightarrow 2\}$ mapping with ordering $\{p^h_1,p_3,p_4,p_2,p^h_5\}$; 
\end{enumerate}

\item $s_{hi} < s_{34}$ and $s_{hi}=s_{35}$,
\begin{enumerate}
\item $s_{12}s_{45} > s_{14}s_{25}$: $\{5\rightarrow 2\}$ mapping with ordering $\{p^h_1,p_3,p_4,p_2,p^h_5\}$; 
\item $s_{12}s_{45} \le s_{14}s_{25}$: $\{5\rightarrow 2\}$ mapping with ordering $\{p^h_1,p_2,p_3,p_4,p^h_5\}$; 
\end{enumerate}

\item $s_{hi} < s_{34}$ and $s_{hi}=s_{45}$,
\begin{enumerate}
\item $s_{12}s_{35} > s_{13}s_{25}$: $\{5\rightarrow 2\}$ mapping with ordering $\{p^h_1,p_3,p_2,p_4,p^h_5\}$;
\item $s_{12}s_{35} \le s_{13}s_{25}$: $\{5\rightarrow 2\}$ mapping with ordering $\{p^h_1,p_2,p_3,p_4,p^h_5\}$. 
\end{enumerate}
\end{enumerate}

\item The evaluation of the $\wt{X}_5^0$ antenna functions is done with the unmapped momenta as usual. The reduced matrix element is evaluated on the mapped momenta corresponding to the appropriate mapping.
\end{enumerate}
Also in this case one can easily verify that the 12 sectors are disjoint and cover the entire phase space.

\subsubsection{Gluon emission between multiple dipoles}

An example of unresolved factor corresponding to a sum of multiple colour connections is given by matrix elements for the scattering of one gluon and two same-flavour quark-antiquark pairs. In this case, a subleading-colour contribution exists where the gluon is colour-connected to all quarks, since they share a unique fermionic line in the squared amplitude. An ambiguity then arise in reconstructing the correct hard momenta when the gluon becomes collinear to a quark (antiquark). Such configuration is addressed at N$^3$LO with the $C_5^0(q_1^h,g_2,q_3,\bar{q}_4,\bar{q}_5^h)$ antenna function, which contains singular terms for all possible $g\parallel q(\bar{q})$ collinear limits and hence requires the definition of dedicated phase-space sectors for the assignment of a momentum mapping. By construction a single $C_5^0(q_1^h,g_2,q_3,\bar{q}_4,\bar{q}_5^h)$ antenna function only contains the $q_3\parallel \bar{q}_4 \parallel \bar{q}_5$ triple-collinear limit and not $q_1\parallel \bar{q}_4 \parallel \bar{q}_5$, $q_1\parallel q_3 \parallel \bar{q}_4$ or $q_1\parallel q_3 \parallel \bar{q}_5$. 
We generically denote such scenario with $\bar{X}_5^0(q_1^h,g_2,q_3,\bar{q}_4,
\bar{q}_5^h)$, given by a sum of contributions with colour connections: $(i-2-j)\oplus(3-4-5)$ with $i\ne j\in\{1,3,4,5\}$. 
Two orderings are sufficient to correctly reconstruct the hard momenta in any case. To identify the appropriate choice according to the phase-space point, we follow the procedure below:
\begin{enumerate}
\item For a given set of momentum $\{p^h_1,p_2,p_3,p_4,p^h_5\}$ in $\bar{X}_5^0(q_1^h,g_2,q_3,\bar{q}_4,
\bar{q}_5^h)$, where $p_2$ is the momentum of the gluon, we evaluate $s_{min}=min\{s_{12},s_{23},s_{24},s_{25},s_{345}\}$, with $s_{345}=s_{34}+s_{35}+s_{45}$.

\item The full phase space is separated into 5 disjoint sectors, where one of the two possible ordering is chosen:
\begin{enumerate}
\item $s_{min} = s_{12}$: $\{5\rightarrow 2\}$ mapping with ordering $\{p^h_1,p_2,p_3,p_4,p^h_5\}$;
\item $s_{min} = s_{23}$: $\{5\rightarrow 2\}$ mapping with ordering $\{p^h_1,p_2,p_3,p_4,p^h_5\}$; 
\item $s_{min} = s_{345}$: $\{5\rightarrow 2\}$ mapping with ordering $\{p^h_1,p_2,p_3,p_4,p^h_5\}$; 
\item $s_{min} = s_{24}$: $\{5\rightarrow 2\}$ mapping with ordering $\{p^h_1,p_3,p_4,p_2,p^h_5\}$; 
\item $s_{min} = s_{25}$: $\{5\rightarrow 2\}$ mapping with ordering $\{p^h_1,p_3,p_4,p_2,p^h_5\}$.
\end{enumerate}

\item The evaluation of $\bar{X}_5^0$ is done with the unmapped momenta as usual. The reduced matrix element is evaluated on the mapped momenta corresponding to the appropriate mapping.
\end{enumerate}
The 5 distinct sectors sum up to recover the full phase space.

We note that the algorithmic procedures described above are not unique, the different orderings can be isolated following other strategies. The relevant point is that their application in a numerical implementation is particularly simple and completely eliminates the necessity of sub-antennae. The phase-space decompositions above have been applied to assign an ordered mapping to the $\vardbtilde{A}_5^0$, $\wt{A}_5^0$ and $C_5^0$ antenna functions to produce the results presented in~\cite{Chen:2025kez}.


\section{Analytical and numerical validation}\label{sec:validation}

In this section we demonstrate the correctness of the procedure described in Section~\ref{sec:sectors}. We first present a general argument to demonstrate that the differential cross sections computed with the sector strategy are equivalent to those obtained if one defines dedicated sub-antennae. Then, we present a series of numerical checks we performed to validate our implementation.

\subsection{Proof of equivalence}

We consider the generic contribution to the cross section of an $\ell$-loop $m$-particle (two hard radiators and $m-2$ unresolved partons) of an antenna function within a subtraction term:
\begin{equation}\label{eq:S}
	S = \int_{\text{full}}\d\Phi_n X_{m}^{\ell}(p_a^h,.,p_b^h)F(.,P_a,.,P_b,.),
\end{equation}
where $P_a$ and $P_b$ are the reconstructed momenta obtained by a $m\to 2$ mapping to reabsorb the unresolved emissions and $F$ collects the reduced matrix element (or possibly secondary antenna functions) and the measurement function, which defines a jet algorithm and the fiducial phase space. The integral above is ill-defined for two reasons. First of all, the integrand is in general divergent, but this is not relevant for the sake of the argument below and we can assume that it is either regulated or that it will be subtracted from the equally divergent contribution of a real-emission matrix element. Secondly, in the presence of unordered emissions it is not possible to uniquely assign $P_a$ and $P_b$ to a given antenna function. 

To do so we can either split the full antenna function into sub-antennae and write the contribution in~\eqref{eq:S} as:
\begin{eqnarray}\label{eq:S1}
	S_1 &=& \int_{\text{full}}\d\Phi_n \sum_{i\in \text{sub.ant.}}x_{m,i}^{\ell}(p_a^h,.,p_b^h)F(.,P^{(i)}_a,.,P^{(i)}_b,.),
\end{eqnarray}
where $P^{(i)}_a$ and $P^{(i)}_b$ are obtained via the specific ordered mapping induced by the sub-antenna $x^{\ell}_{m,i}$, or adopt the sector strategy and write the contribution in~\eqref{eq:S} as:
\begin{eqnarray}\label{eq:S2}
	S_2 &=&  \sum_{k\in \text{sec}}\int_{k}\d\Phi_n X_{m}^{\ell}(p_a^h,.,p_b^h)F(.,P^{(k)}_a,.,P^{(k)}_b,.) ,
\end{eqnarray}
where the subscript $k$ on the integral indicates the restriction to a specific sector and $P^{(k)}_a$ and $P^{(k)}_b$ are reconstructed with the ordered mapping appropriate for that sector. In both cases, the mapping correctly behaves in all the infrared configurations by construction.

To show that the $S_1$ and $S_2$ are the same, we can separate the phase space into sectors in $S_1$ and split the full antenna into sub-antennae in $S_2$. Indeed the fact that the sum over all sectors gives the full phase space, as well as that the sum of all sub-antennae gives the original antenna functions hold regardless of the employed strategy. We then obtain:
\begin{eqnarray}
	S_1 &=&  \sum_{k\in \text{sec}}\int_{k}\d\Phi_n \sum_{i\in \text{sub.ant.}}x_{m,i}^{\ell}(p_a^h,.,p_b^h)F(.,P^{(i)}_a,.,P^{(i)}_b,.)\nn \\
	       &=&\sum_{\substack{k\in \text{sec}}}\int_k\d\Phi_{X_{m}} \sum_{i\in\text{sub.ant.}}x_{m,i}^{\ell}(p_a^h,.,p_b^h)\int_{\text{full}}\d\Phi_{n-m+2}F(.,P^{(i)}_a,.,P^{(i)}_b,.),
\end{eqnarray}
and
\begin{eqnarray}
	S_2 &=& \sum_{k\in \text{sec}}\int_{k}\d\Phi_n \sum_{i\in \text{sub.ant.}}x_{m,i}^{\ell}(p_a^h,.,p_b^h)F(.,P^{(k)}_a,.,P^{(k)}_b,.)\nn \\
	&=&\sum_{\substack{k\in \text{sec} }}\int_k\d\Phi_{X_{m}}\sum_{i\in \text{sub.ant.}} x_{m,i}^{\ell}(p_a^h,.,p_b^h)\int_{\text{full}}\d\Phi_{n-m+2}F(.,P^{(k)}_a,.,P^{(k)}_b,.),
\end{eqnarray}
where in the last step of both equations above we relied on the factorisation of the phase space (eq.\eqref{eq:PSfac}) into an antenna phase space $\d\Phi_{X_{m}}$, for the inclusive integration over the $(m-2)$ unresolved partons, and a reduced phase space $\d \Phi_{n-m+2}$, as described in Section~\ref{sec:mapping}. The reduced phase space is not affected by the division into sectors, which are defined by relations among the Mandelstam invariants of the unmapped momenta living in the antenna phase space only.

The difference of the two approaches then reads
\begin{eqnarray}
	S_1-S_2 &=& \sum_{\substack{k\in \text{sec} }}\int_k\d\Phi_{X_{m}} \sum_{i\in\text{sub.ant.}} x_{m,i}^{\ell}(p_a^h,.,p_b^h)\nn\\
	&\phantom{=}&\hspace{-1cm}\cdot\left[\int_{\text{full}}\d\Phi_{n-m+2}F(.,P^{(i)}_a,.,P^{(i)}_b,.)-\int_{\text{full}}\d\Phi_{n-m+2}F(.,P^{(k)}_a,.,P^{(k)}_b,.)\right]=0.
\end{eqnarray}
The conclusion follows from the fact that the two integrals in the square brackets differ by a simple re-labelling of the mapped momenta $P_a$ and $P_b$, which does not change the result of the integration over the reduced phase space. In general, the cancellation does not happen locally across the phase space. Indeed two different mappings reconstruct different hard momenta, especially away from infrared limits.  Nevertheless, the final result after integration does not depend on the specific choice of the $m\to 2$ mapping, and in particular on the fact that it is induced by a sub-antenna or by a phase-space sector, thanks to the factorization of the reduced phase space. The equivalence of the results obtained with the two methods is then established. 

\begin{figure}[h]
	\centering
	\includegraphics[width=0.49\columnwidth]{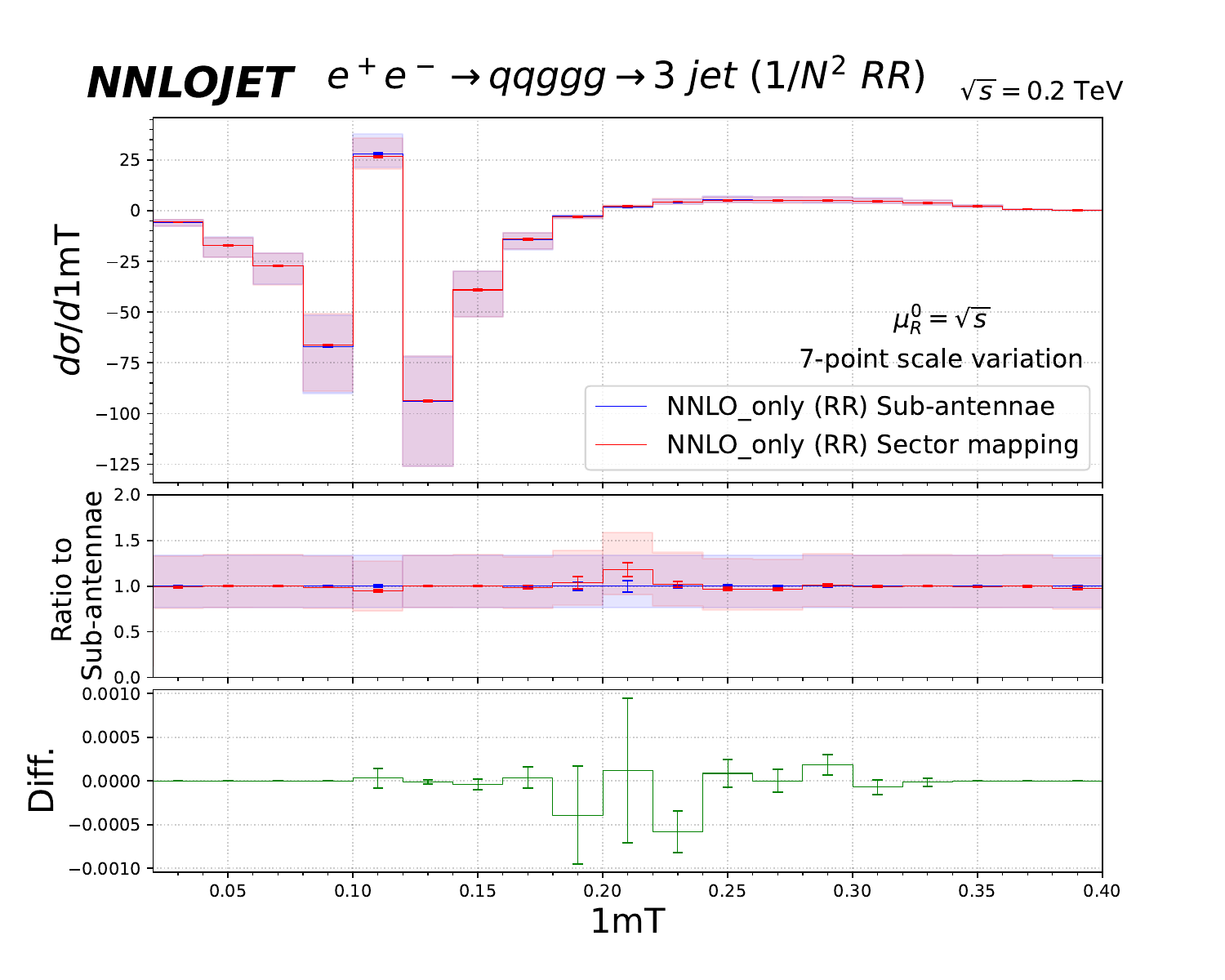}
	\includegraphics[width=0.49\columnwidth]{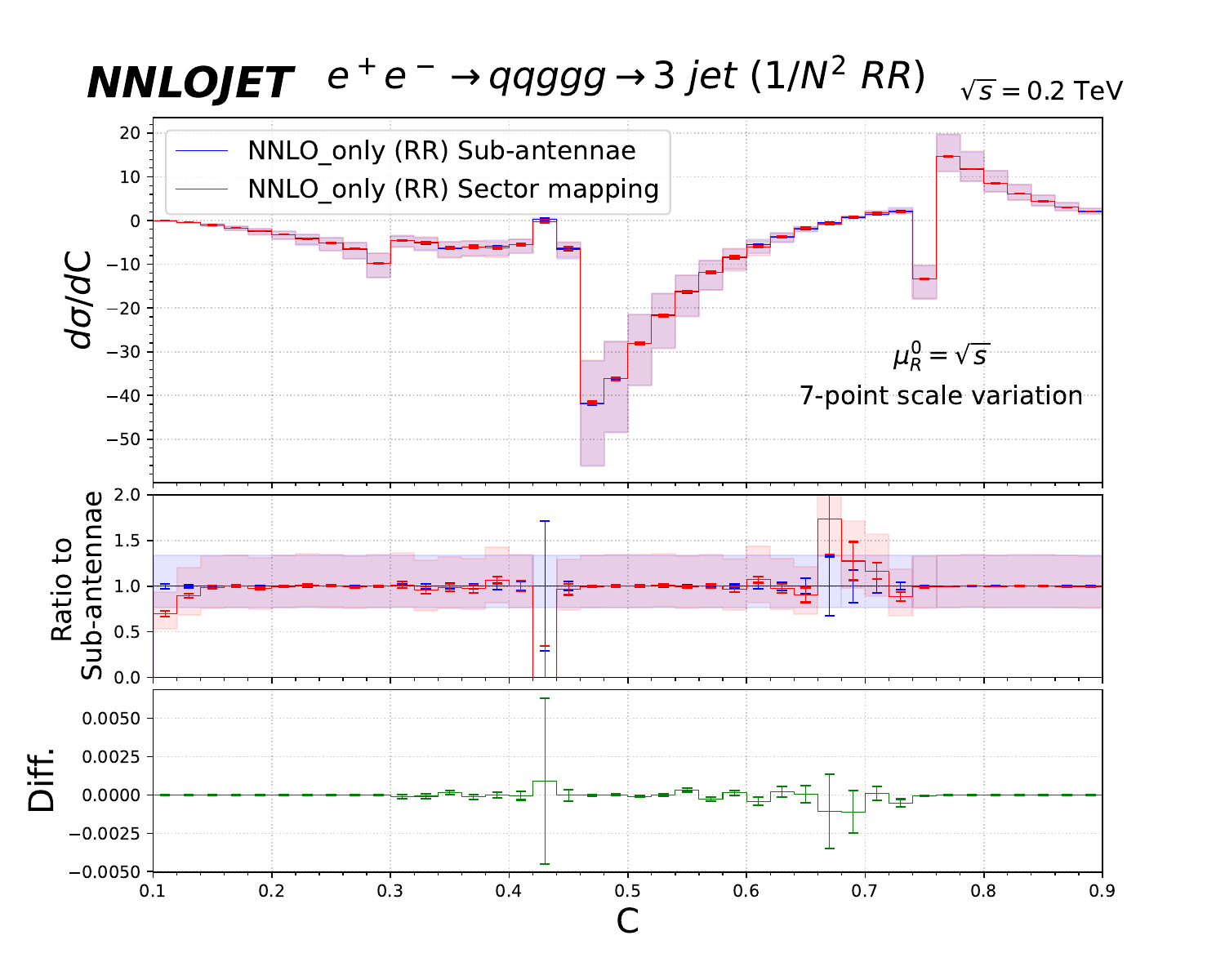}\\
	\includegraphics[width=0.49\columnwidth]{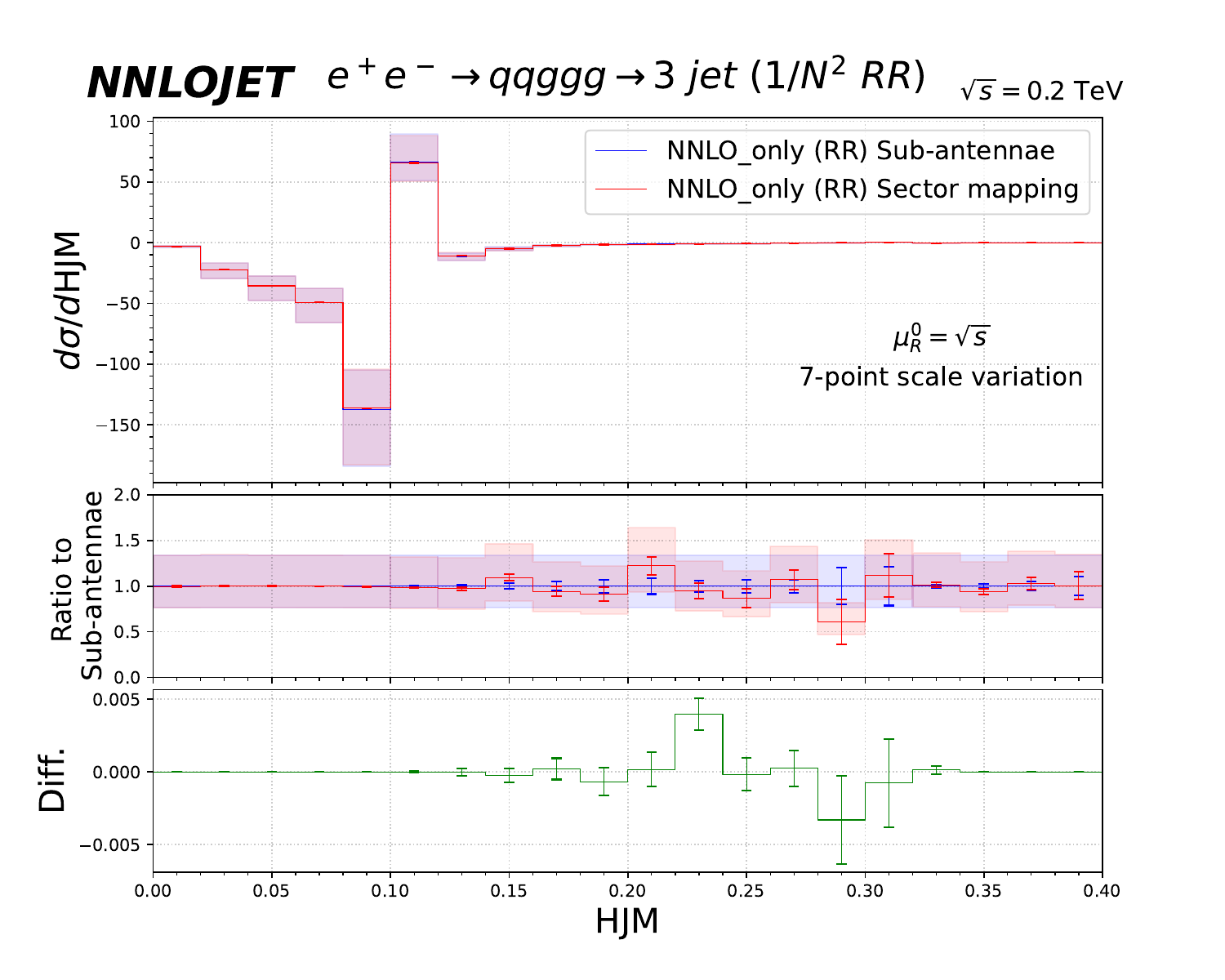}
	\includegraphics[width=0.49\columnwidth]{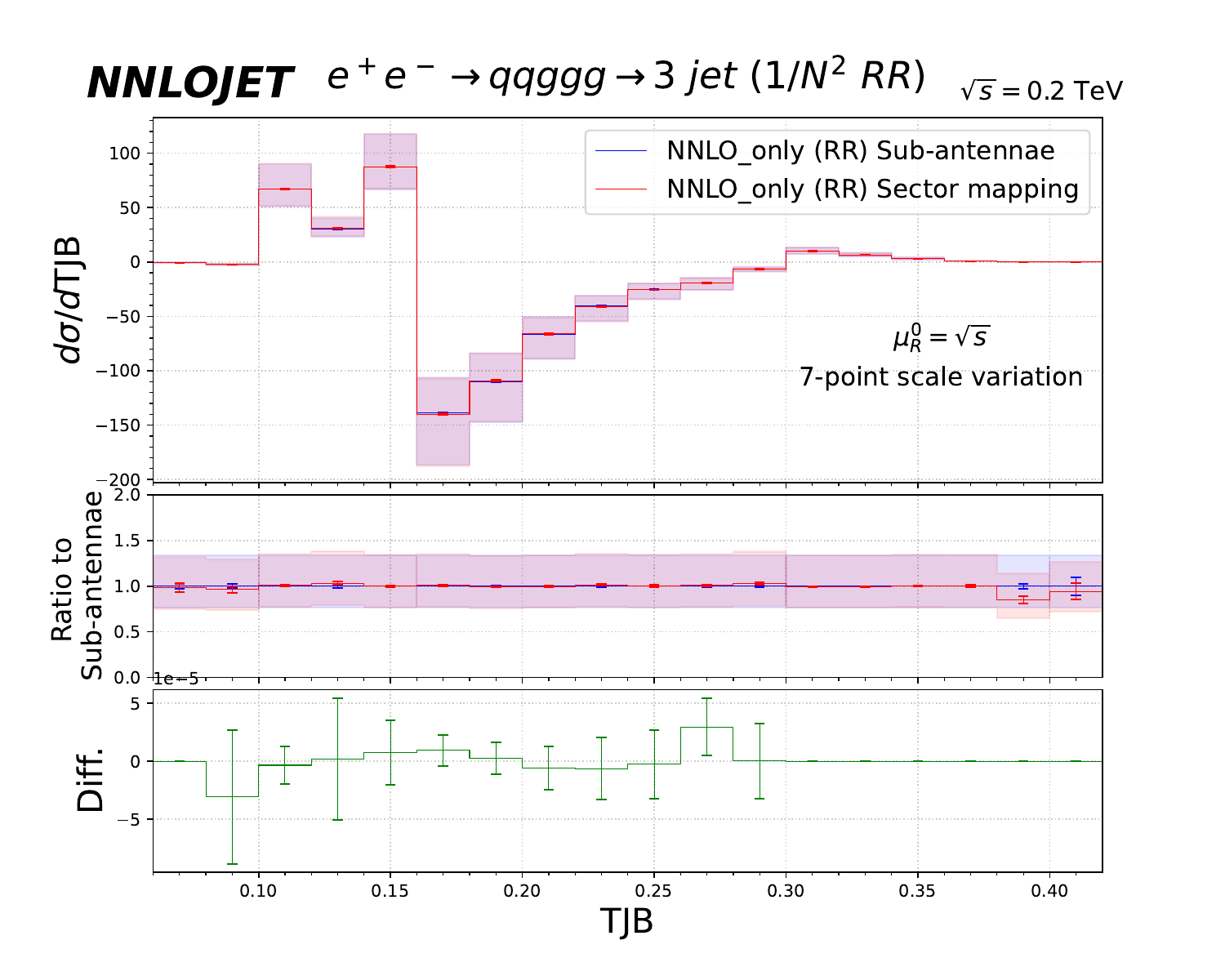}
	\caption{Comparison for selected event shapes in electron-positron annihilation between the original antenna subtraction implementation relying on the decomposition into sub-antennae and the new one employing the phase-space sector-based strategy to identify the proper momentum mapping. Subtracted double-real contribution to the $N_c^{-2}$ colour factor (using the $\tilde{A}_4^0$ antenna function), as defined in~\cite{Gehrmann-DeRidder:2007foh}.}
	\label{fig:Btt}
\end{figure}

\subsection{Numerical tests}
 
 We performed a series of numerical tests, which are reported below. At NNLO we verified that cross-section level results obtained by implementing the sector strategy described in this paper are in complete agreement with the original implementation of antenna subtraction relying on sub-antennae.  At N$^3$LO, where the only available implementation is the one based on phase-space sectors, and for selected configurations at NNLO, we show that the subtraction terms which rely on the procedure described in this paper for the assignment of momentum mappings are fully capable of subtracting the infrared divergences of the real-emission matrix elements locally across the phase space. This validation is done following the method described for example in~\cite{Glover:2010kwr,Gehrmann:2023dxm}.
 
\subsubsection{Double-unresolved sectors}

As discussed in Section~\ref{sec:2unsectors}, the only case to be addressed at NNLO is the emission of two abelian gluons and the affected antenna functions are $\wt{A}_4^0$, $D_{4,c}^0$ and $F_{4,b}^0$, plus the analogous antenna functions constructed via the designer antenna approach and higher-loop counterparts of these. 

To test the behaviour of the $\wt{A}_4^0$ and $D_{4,c}^0$ antenna functions we computed the subtracted double-real contribution to the NNLO correction to event shapes in electron-positron annihilation. As described in detail in~\cite{Gehrmann-DeRidder:2007foh}, the $\wt{A}_4^0$ and $D_{4,c}^0$  antenna functions enter the double-real subtraction terms for the $N_c^{-2}$ and $N_c^{2}$ colour factors respectively. See~\cite{Gehrmann-DeRidder:2007foh} for the specific conventions on the definition of the colour factors. We therefore consider the process $e^+e^-\to jjj$ and four standard event shapes: one-minus-thrust ($1mT$), C-parameter ($C$), heavy jet mass ($HJM$) and total jet broadening ($TJB$), and compute the NNLO correction to them with the original antenna-subtraction implementation relying on the definition of sub-antennae and with the newly proposed phase-space sector approach.  By focusing on the specific colour factors where the antenna functions above enter, rather than on the full result, we can better resolve a potential disagreement.
\begin{figure}[h]
	\centering
	\includegraphics[width=0.49\columnwidth]{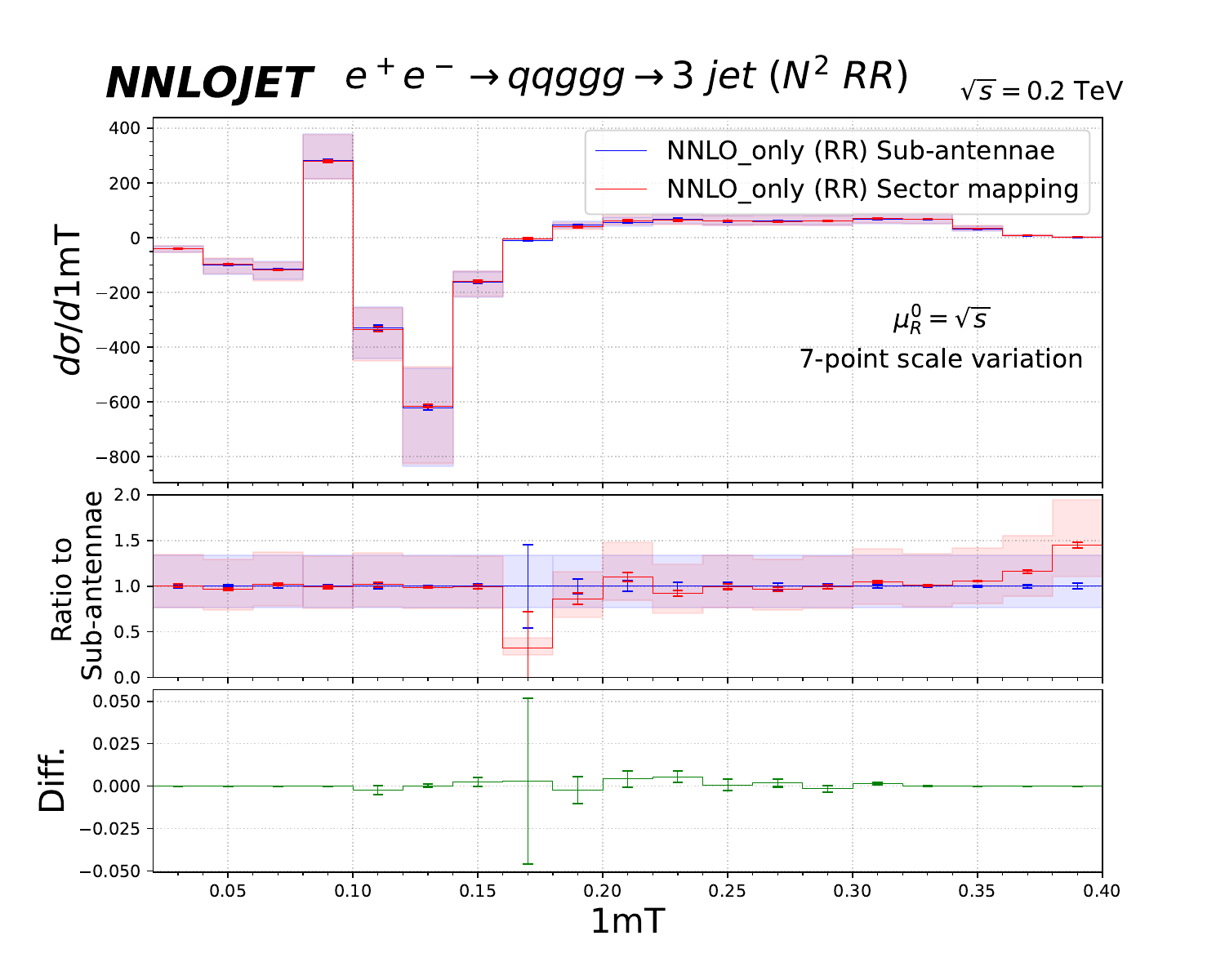}
	\includegraphics[width=0.49\columnwidth]{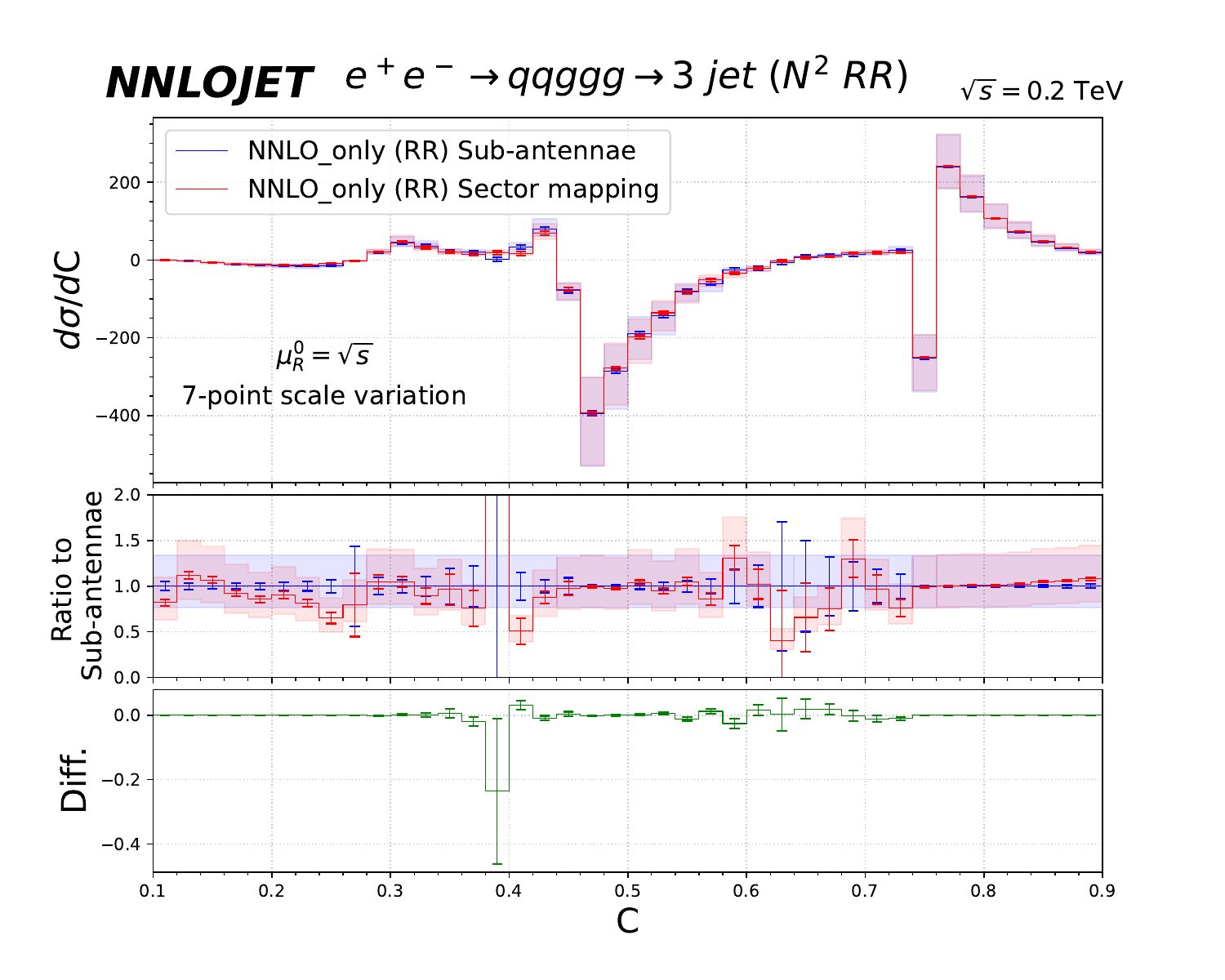}\\
	\includegraphics[width=0.49\columnwidth]{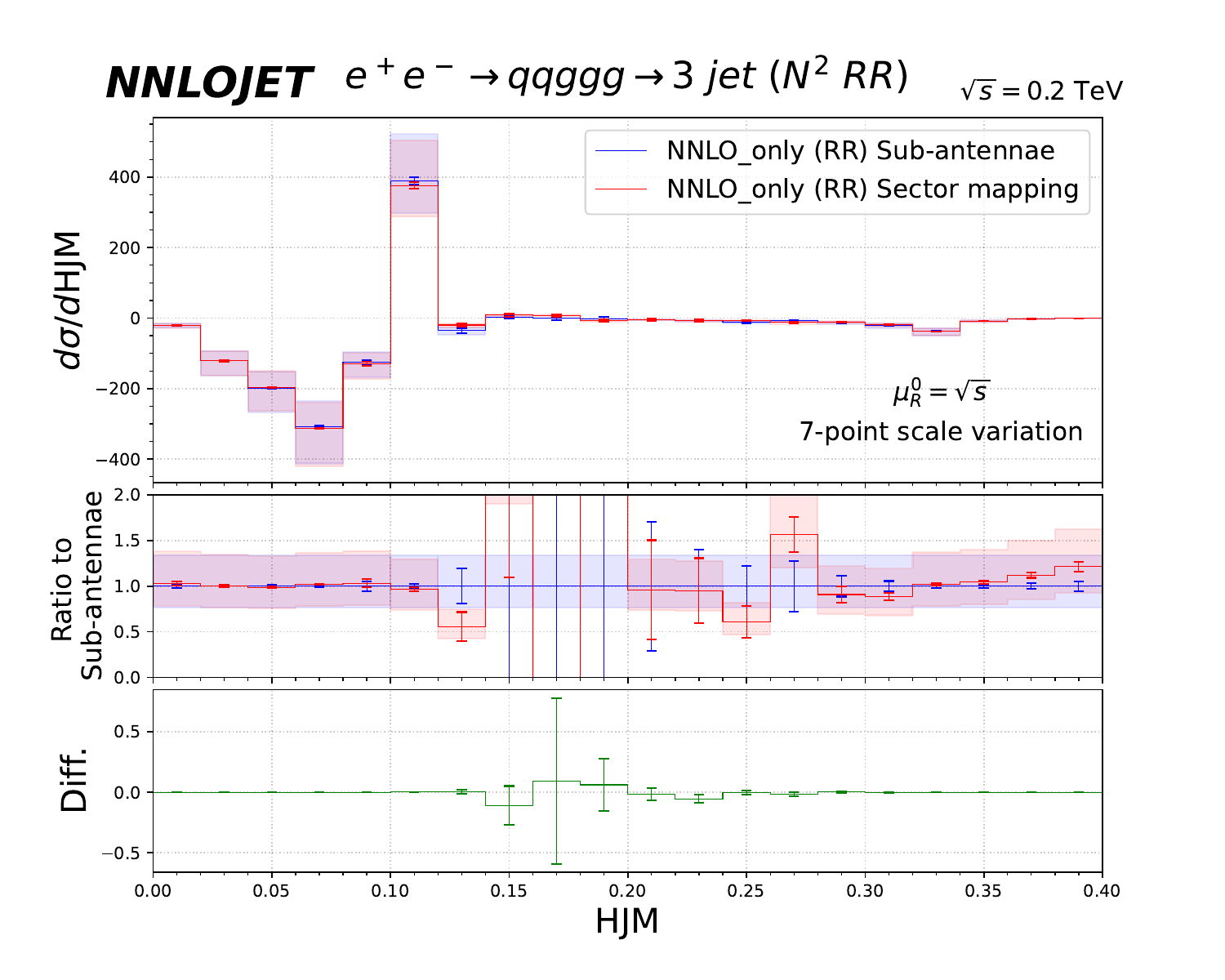}
	\includegraphics[width=0.49\columnwidth]{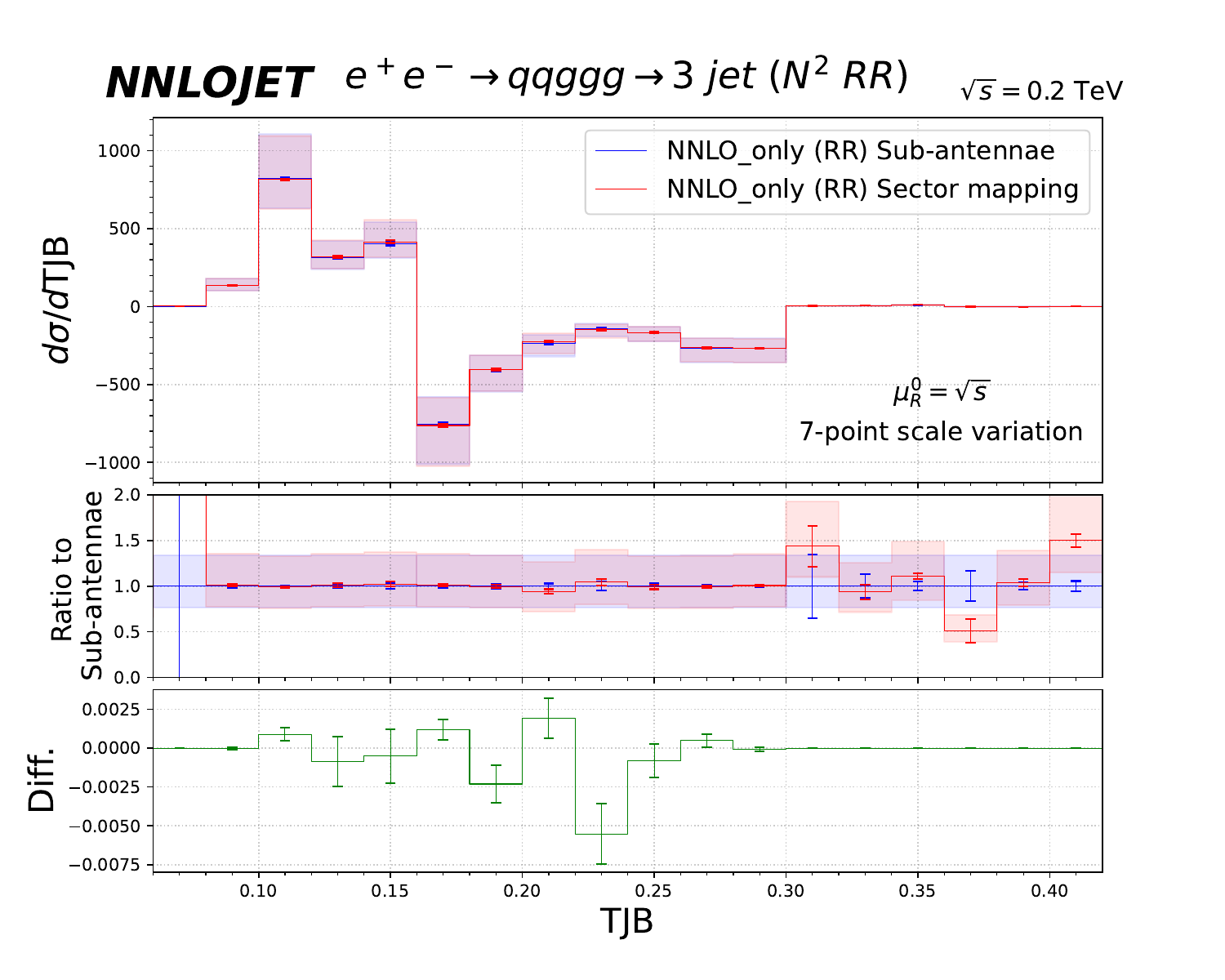}
	\caption{Comparison for selected event shapes in electron-positron annihilation between the original antenna subtraction implementation relying on the decomposition into sub-antennae and the new one employing the phase-space sector-based strategy to identify the proper momentum mapping. Subtracted double-real contribution to the $N_c^{2}$ colour factor (using the $D_{4,c}^0$ antenna function), as defined in~\cite{Gehrmann-DeRidder:2007foh}.}
	\label{fig:B}
\end{figure}
We present the results of the comparison in Figure~\ref{fig:Btt} for the $N_c^{-2}$ colour factor and in Figure~\ref{fig:B} for the $N_c^{2}$ colour factor. We compute the differential cross section (NNLO coefficient, RR contribution)  obtained with the two approaches and observe complete agreement. Large Monte Carlo error bars in the ratio-plots are simply due to the distributions approaching numerically small values. In the last panel of each plot, we also present the result of the Monte Carlo integration of the event-by-event difference between the two implementations. We observe that the result is fully compatible with zero, with larger Monte Carlo errors in the bins where the differential distributions vanish, further supporting the agreement. 
We remark that, although the numerical validation above refers to a process with final-state QCD radiation only, the affected antenna functions and final-state momentum mappings will appear in any calculation performed with antenna subtraction, including the ones for hadronic collisions. 

To probe such a process, we consider the leading-colour contribution to di-jet production at hadron colliders $pp\rightarrow jj$. The double-real subtraction term for such a process features the $F_{4,b}^0$ antenna function~\cite{Glover:2010kwr}. In this case, given that cross section calculations are computationally costly, we compare the performance of the original sub-antenna implementation and the phase-space sector strategy by inspecting the quality of the cancellation of infrared divergences between the double-real matrix element and the subtraction term. 

We consider gluon-gluon final-state single- and double-collinear limits, where a wrongly-ordered mapping would fail to reconstruct the correct hard momenta.  The numerical checks we perform~\cite{Gehrmann:2023dxm} consists in generating phase-space points close to the collinear limits, with different infrared depths, parametrised by 
\begin{equation}
	x = \dfrac{s_{i_1\dots i_n}}{\hat{s}},
\end{equation}
where $s_{i_1\dots i_n}$ is the invariant mass squared of the $n$ (here $n=2$) collinear partons and $\hat{s}$ is the partonic centre-of-mass energy squared. The smaller $x$ is, the closer the events are to the exact collinear limit. In case of unresolved limits involving multiple small invariants, each one is suppressed by a factor of $x$ with respect to the hard scale. We can then study point-by-point the cancellation between the double-real matrix element $M_{RR}$ and the subtraction term $S_{RR}$ by computing the variable:
\begin{equation}\label{eq:tRR}
	t_{RR} = \log_{10}\left(\left|1-\dfrac{M_{RR}}{S_{RR}}\right|\right),
\end{equation}
which gives the number of digits of cancellation up to its sign.

\begin{figure}[t]
 \centering
\includegraphics[width=0.49\columnwidth]{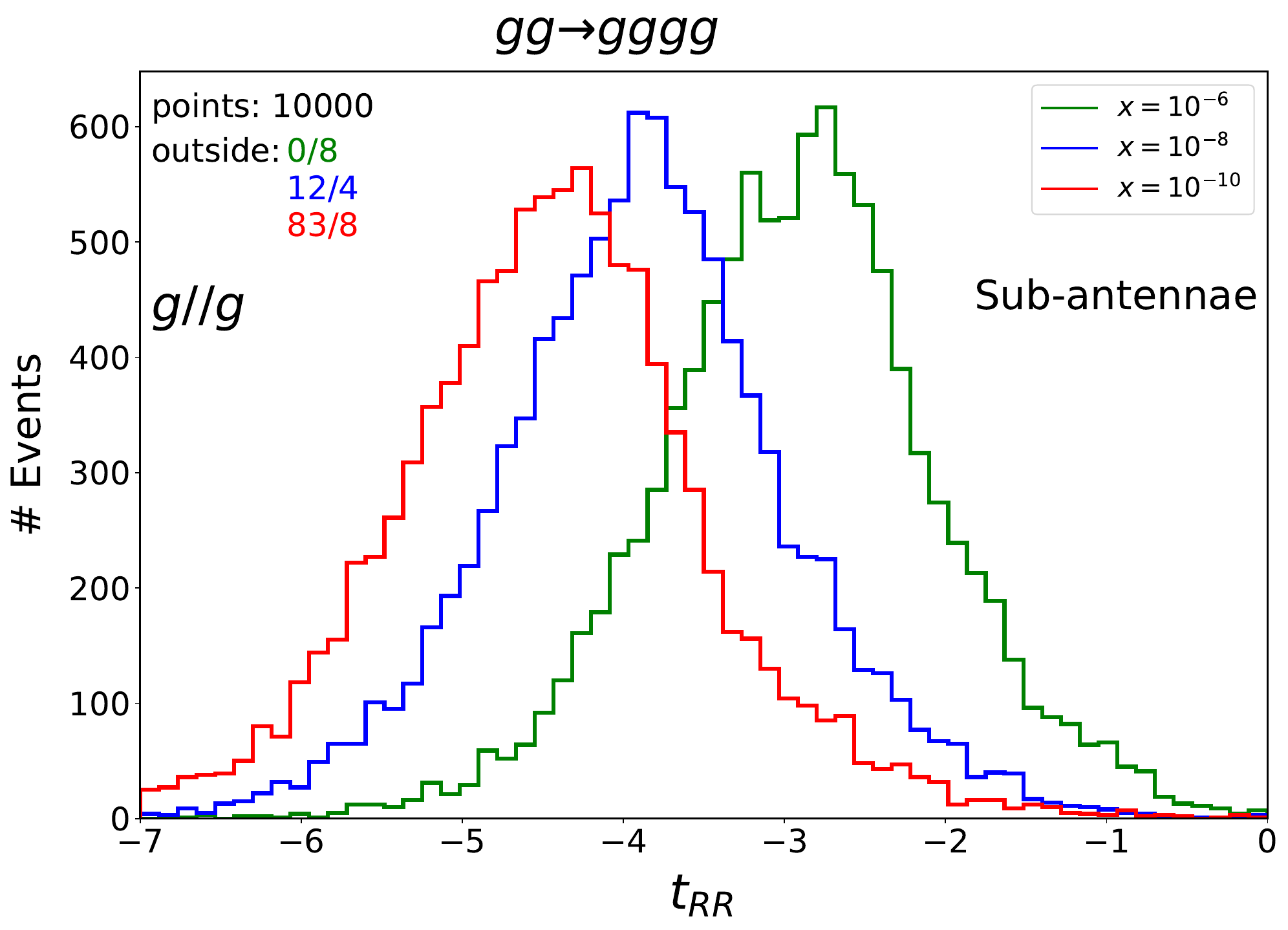}
\includegraphics[width=0.49\columnwidth]{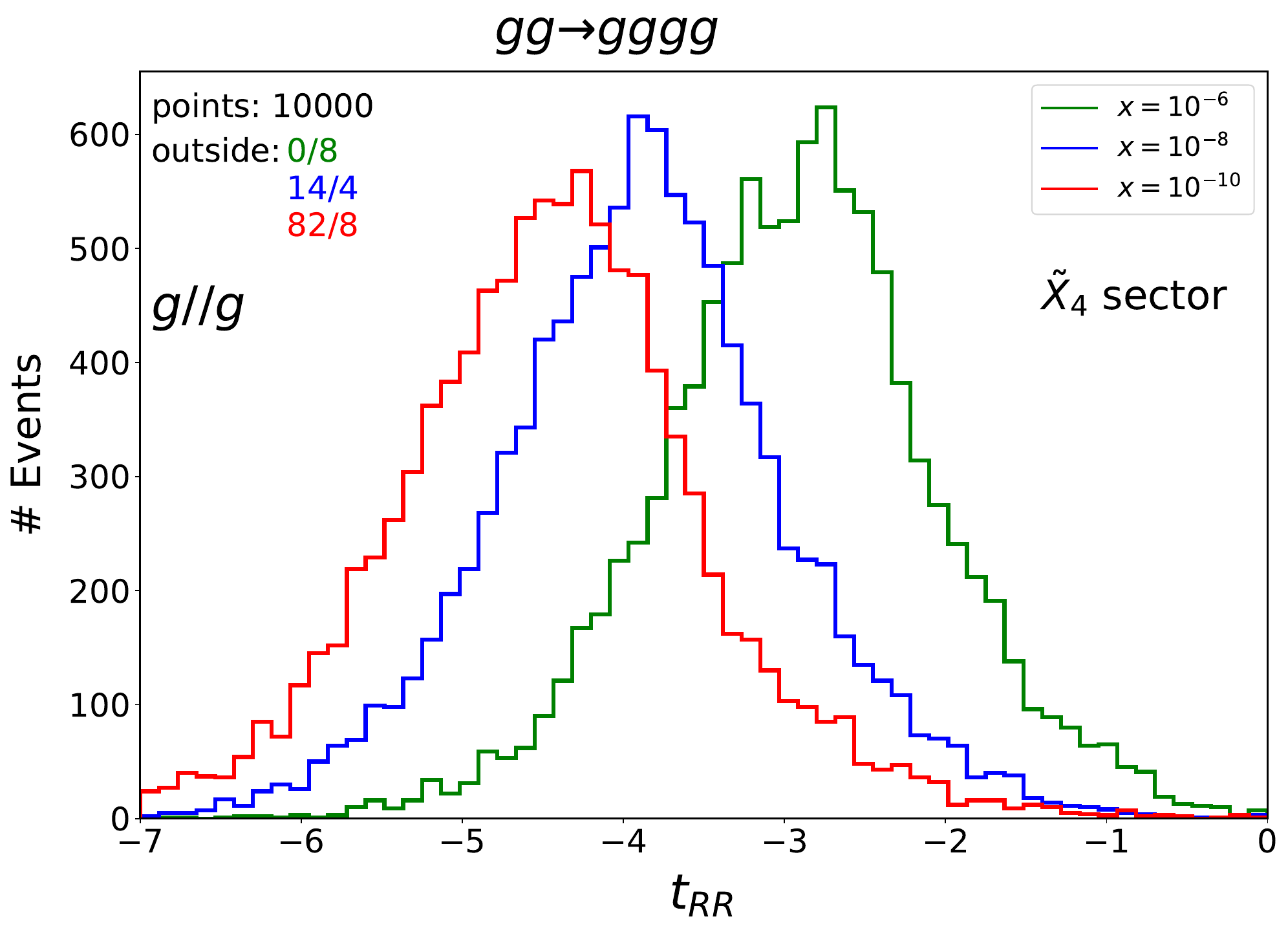}\\
\includegraphics[width=0.49\columnwidth]{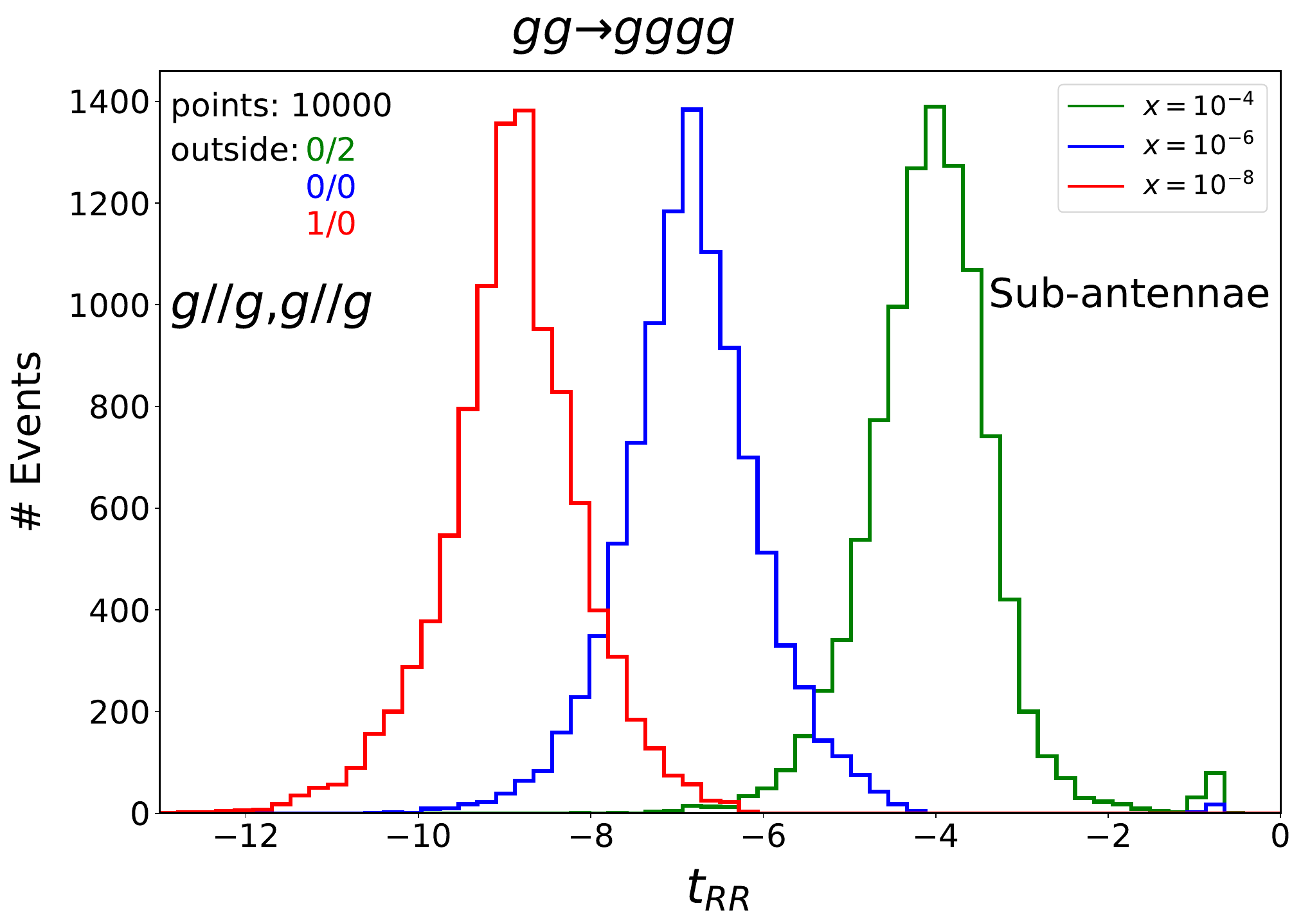}
\includegraphics[width=0.49\columnwidth]{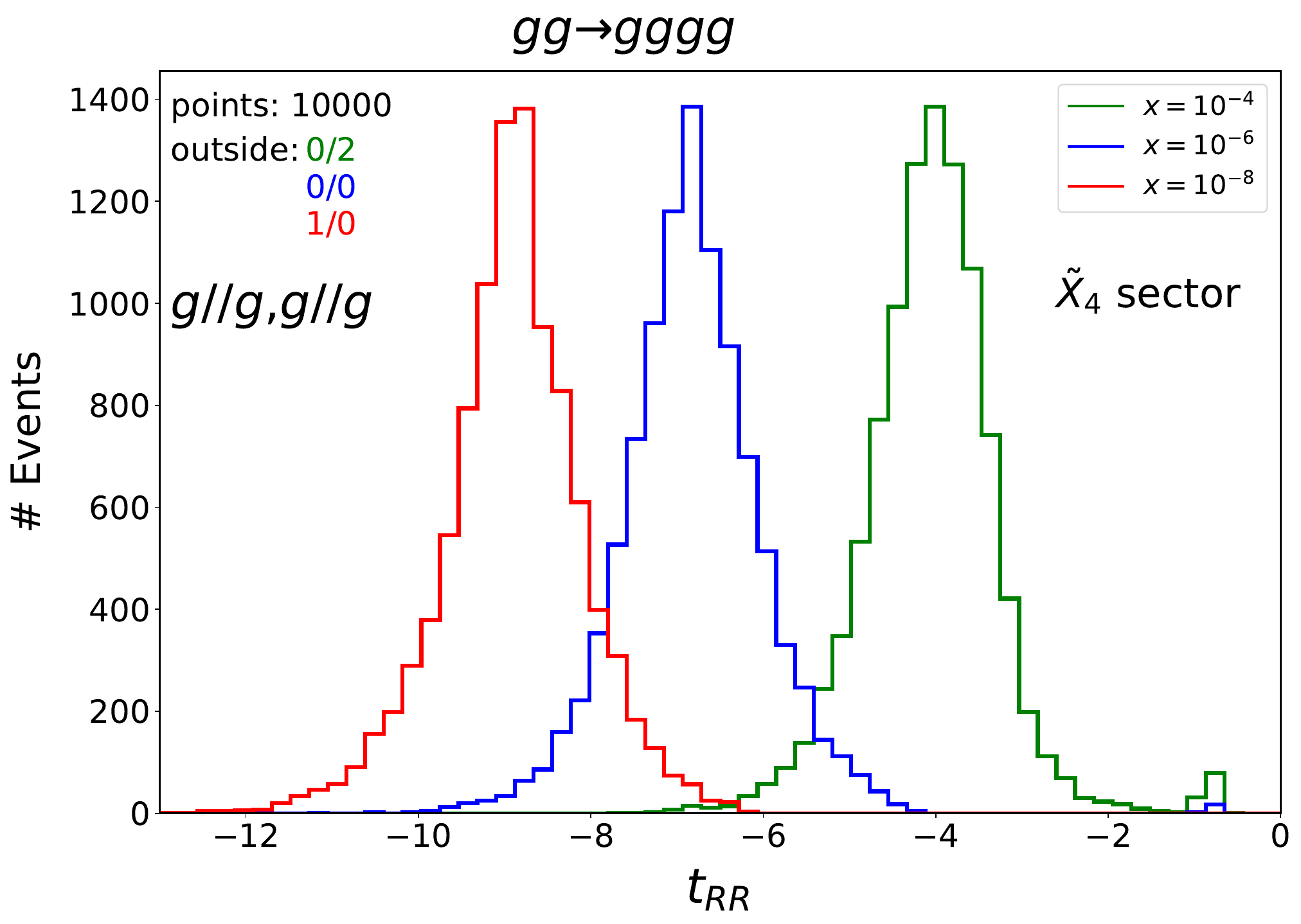}
\caption{Distribution of the $t_{RR}$ variable for the leading-colour contribution to the $gg\to gggg$ sub-process, using the $F_{4,b}^0$ antenna function, as a validation of the newly proposed implementation against the original one. The frames refer to single-collinear limits (upper row) and double-collinear ones (lower row), for the results obtained with the original implementation relying on sub-antenna functions (left) and with sector-based approach (right).}
\label{fig:F40b}
\end{figure}

In Figure~\ref{fig:F40b} we compute the distribution of $t_{RR}$ for 10000 points at three different infrared depths ($x=10^{-6}$, $10^{-8}$ and $10^{-10}$ for single-unresolved and $x=10^{-4}$, $10^{-6}$ and $10^{-8}$ for double-unresolved) using the standard subtraction term (left) and a new one implementing the sector-based momentum mapping (right). The `$n_1/n_2$' numbers labelled as  `outside' in the plots indicate the number of events for which $t_{RR}$ falls to the left ($n_1$) or to the right ($n_2$) of the shown range.
The phase-space points are the same for the two cases. The $t_{RR}$ distributions are basically the same, namely the new approach yields no visible degradation in the reconstruction of the hard momenta, and so in the infrared cancellation, with respect to the original approach. We note that there is no improvement either, but this is not surprising: the advantages of the approach we discuss in this paper lie in the simplicity of its application and scalability at higher orders, not in an improved infrared convergence, given that the actual momentum mapping (antenna mapping) is kept the same as before. 

\begin{figure}[h]
 \centering
\includegraphics[width=0.49\columnwidth]{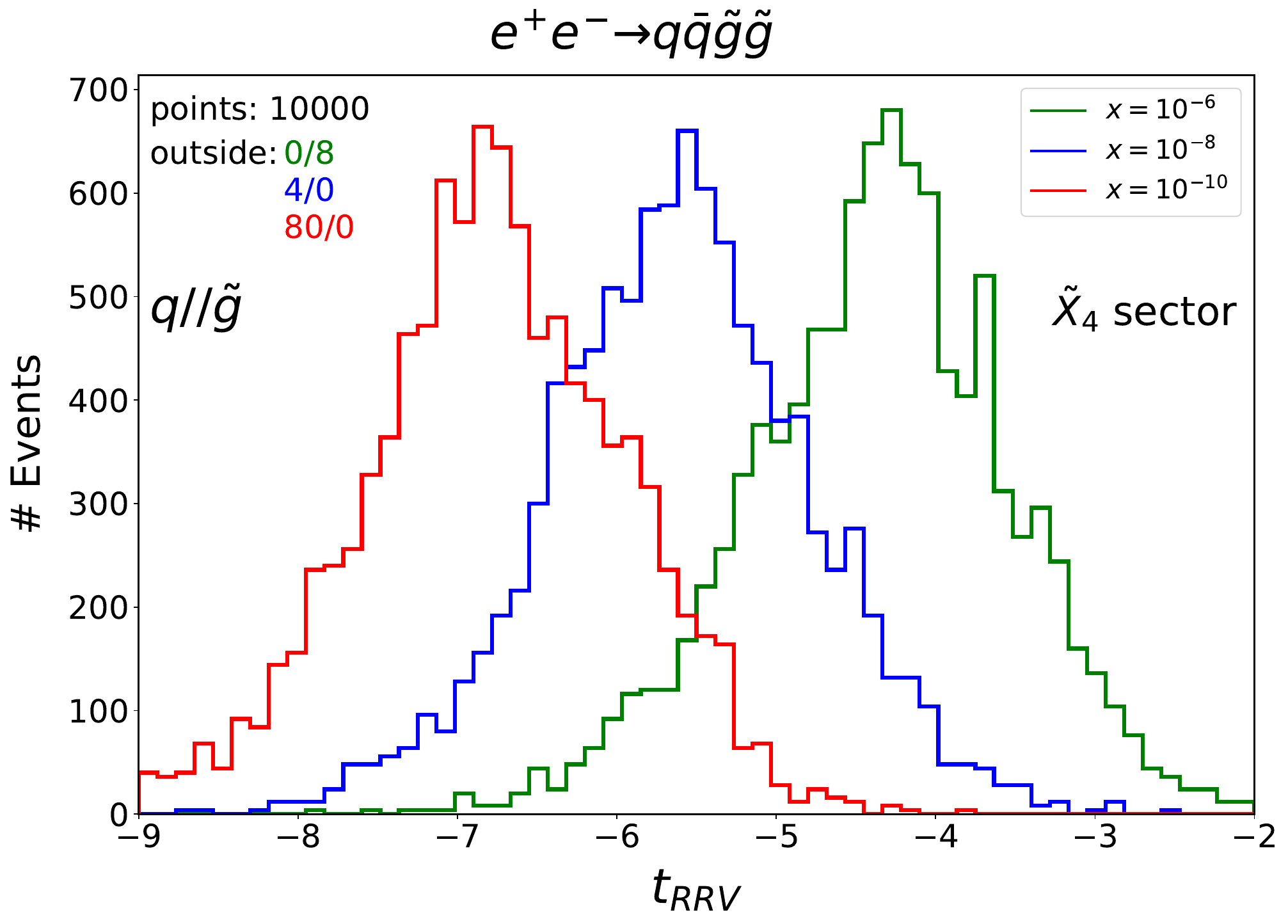}
\includegraphics[width=0.49\columnwidth]{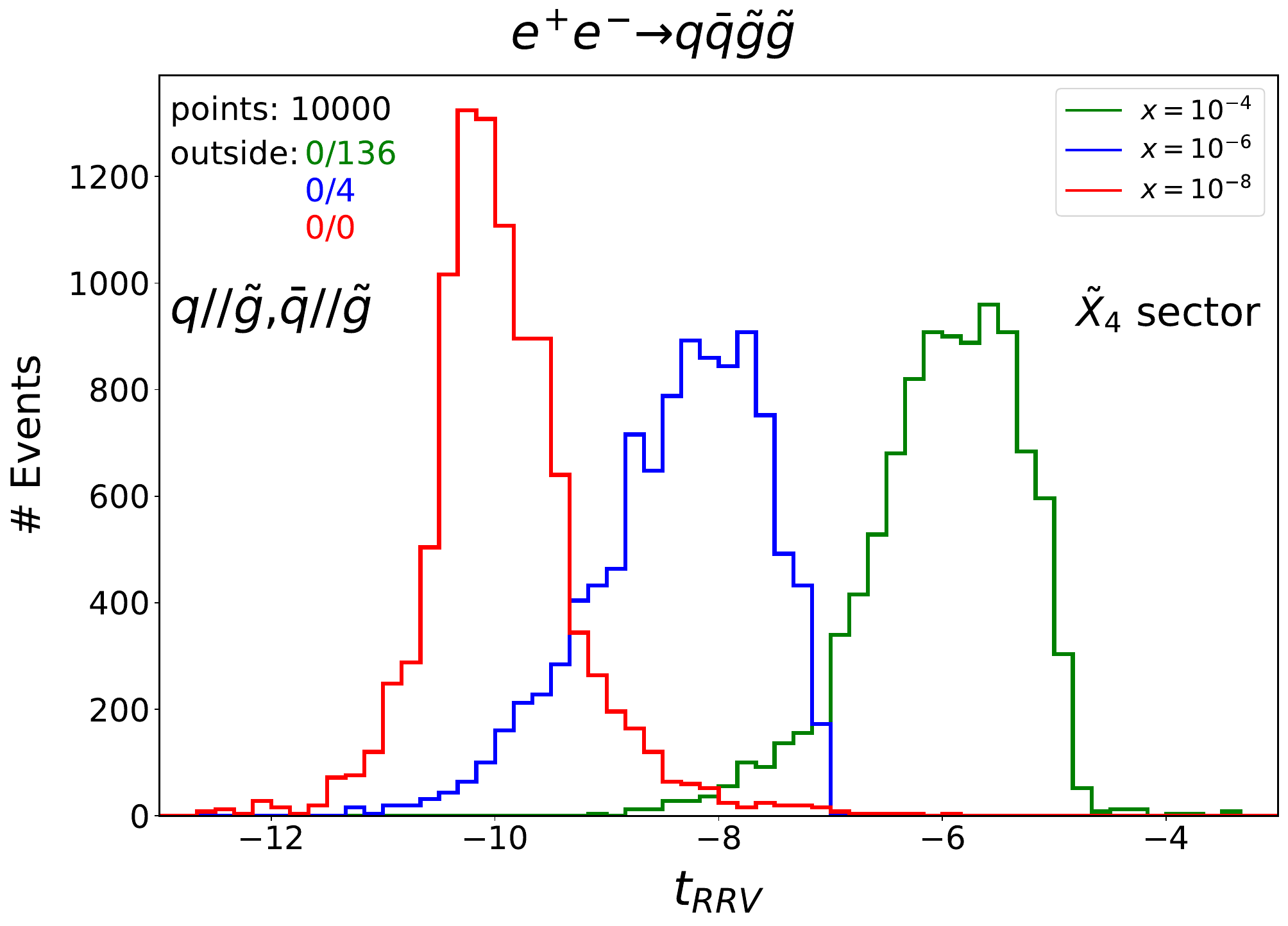}
\caption{Distribution of the $t_{RRV}$ variable for the subleading-colour contribution to the $e^+ e^-\to q\bar{q}gg$ sub-process, using the $\wt{A}_{4}^1$ antenna function. The frames refer to single-collinear limits (left) and double-collinear ones (right).}
\label{fig:At41}
\end{figure}

Finally, we show an analogous numerical test for the one-loop $\wt{A}_4^1$ antenna function, which has been used in~\cite{Chen:2025kez} as part of the double-real-virtual subtraction term for calculation of the N$^3$LO correction to jet production in electron-positron annihilation. In Figure~\ref{fig:At41}, we present the $t_{RRV}$ distribution, which is defined analogously to $t_{RR}$ in~\eqref{eq:tRR}, for a quark-gluon collinear limit at subleading-colour. 
We clearly observe that the deeper the collinear limit is probed, the more digits are cancelled between the matrix element and the subtraction term, indicating that the momentum mappings are working correctly. In this case there is no comparison with the sub-antenna approach because, given the success of the phase-space sector strategy, there was no need to embark in a partial-fractioning procedure of the $\wt{A}_4^1$ antenna function for the numerical implementation of~\cite{Chen:2025kez}.

\subsubsection{Triple-unresolved sectors}

\begin{figure}[t]
	\centering
	\includegraphics[width=0.32\columnwidth]{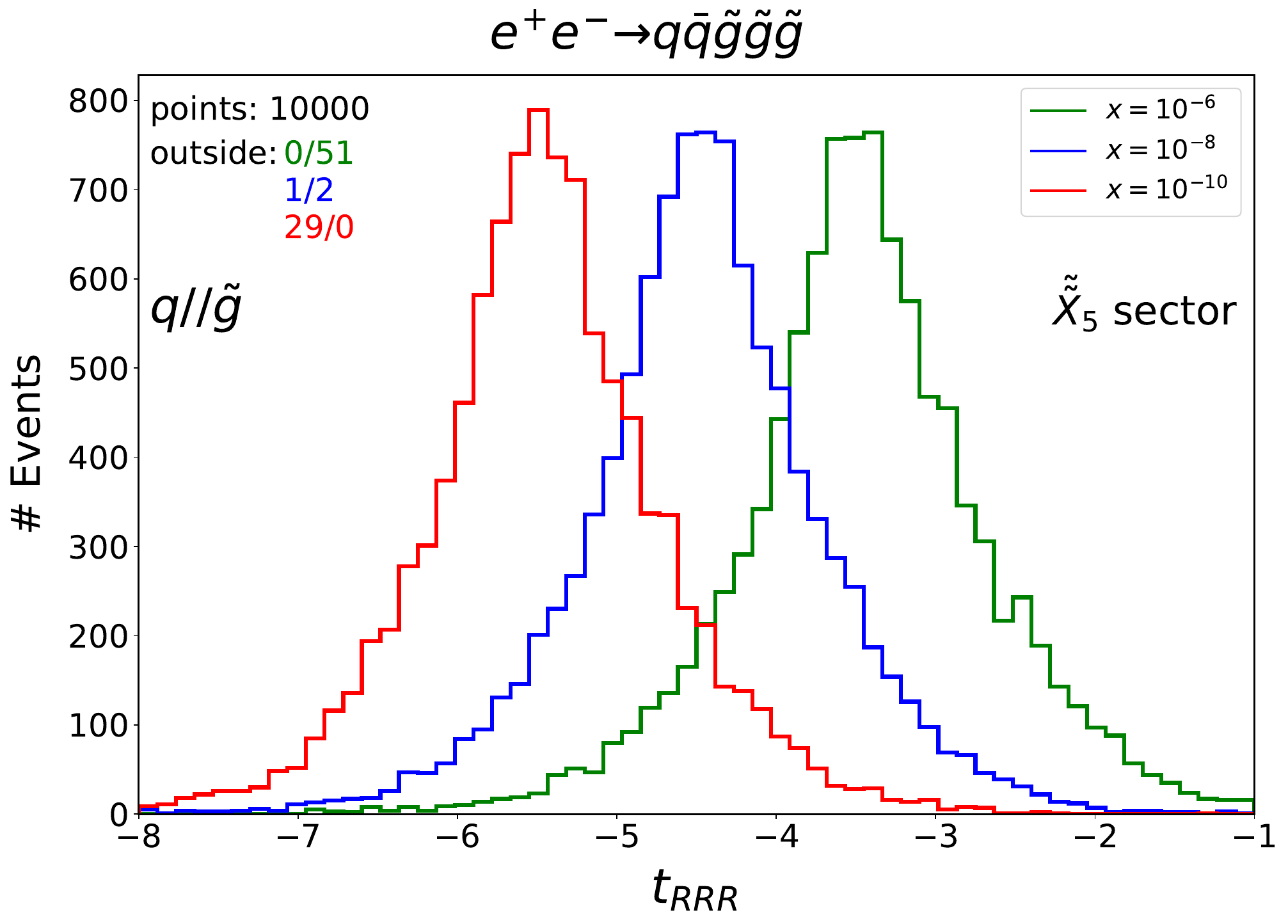}
	\includegraphics[width=0.32\columnwidth]{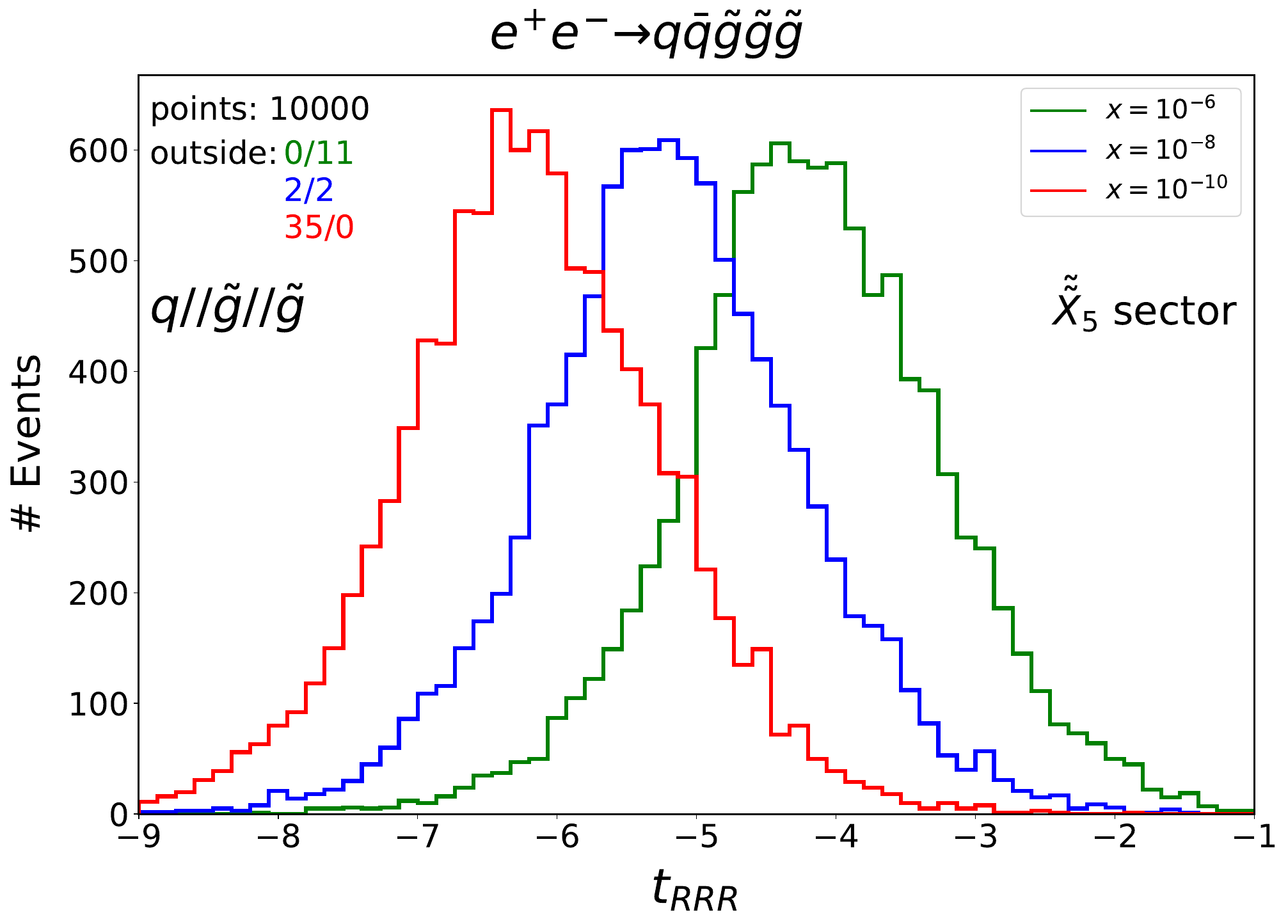}
	\includegraphics[width=0.32\columnwidth]{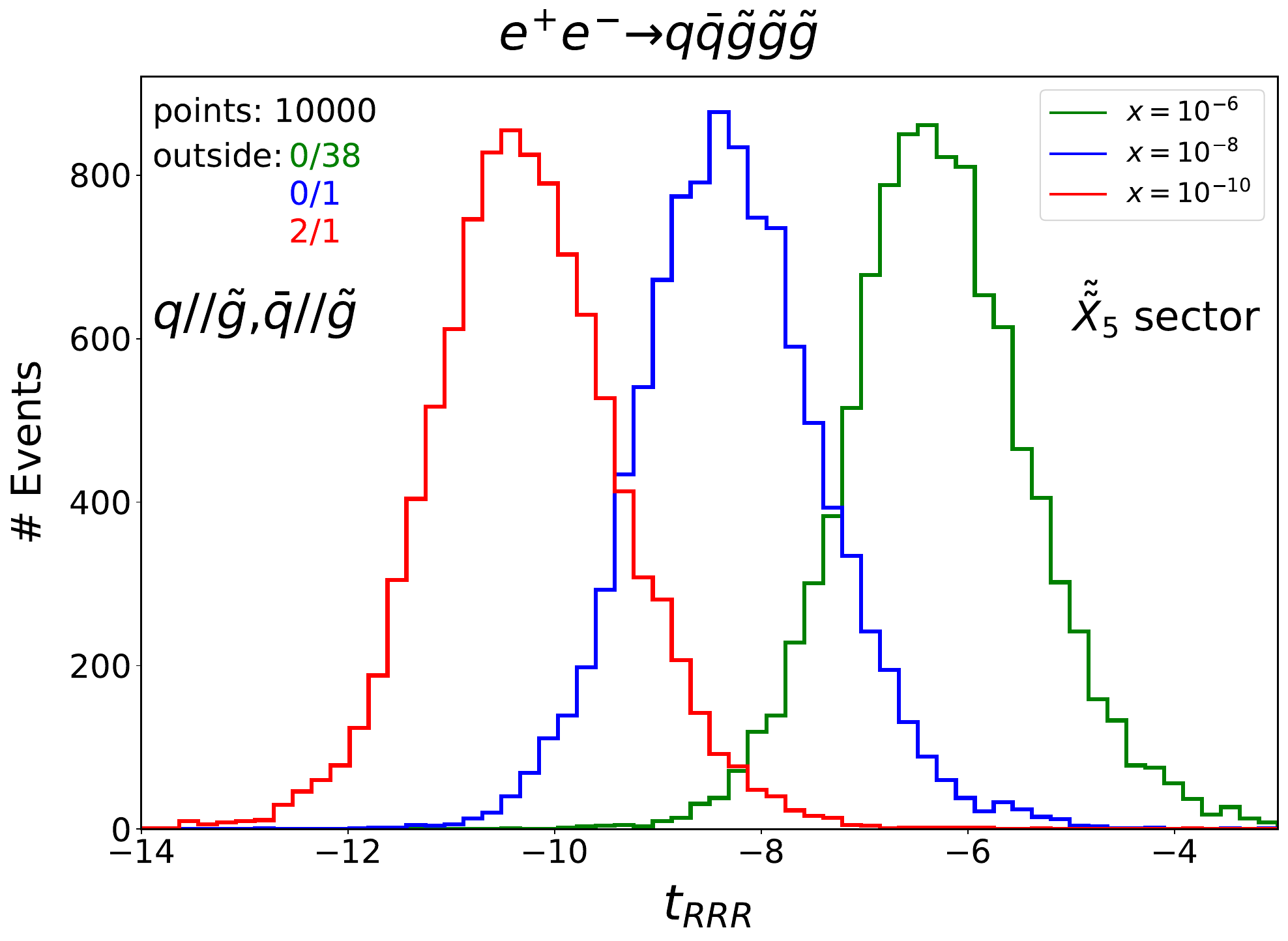}\\
    \includegraphics[width=0.32\columnwidth]{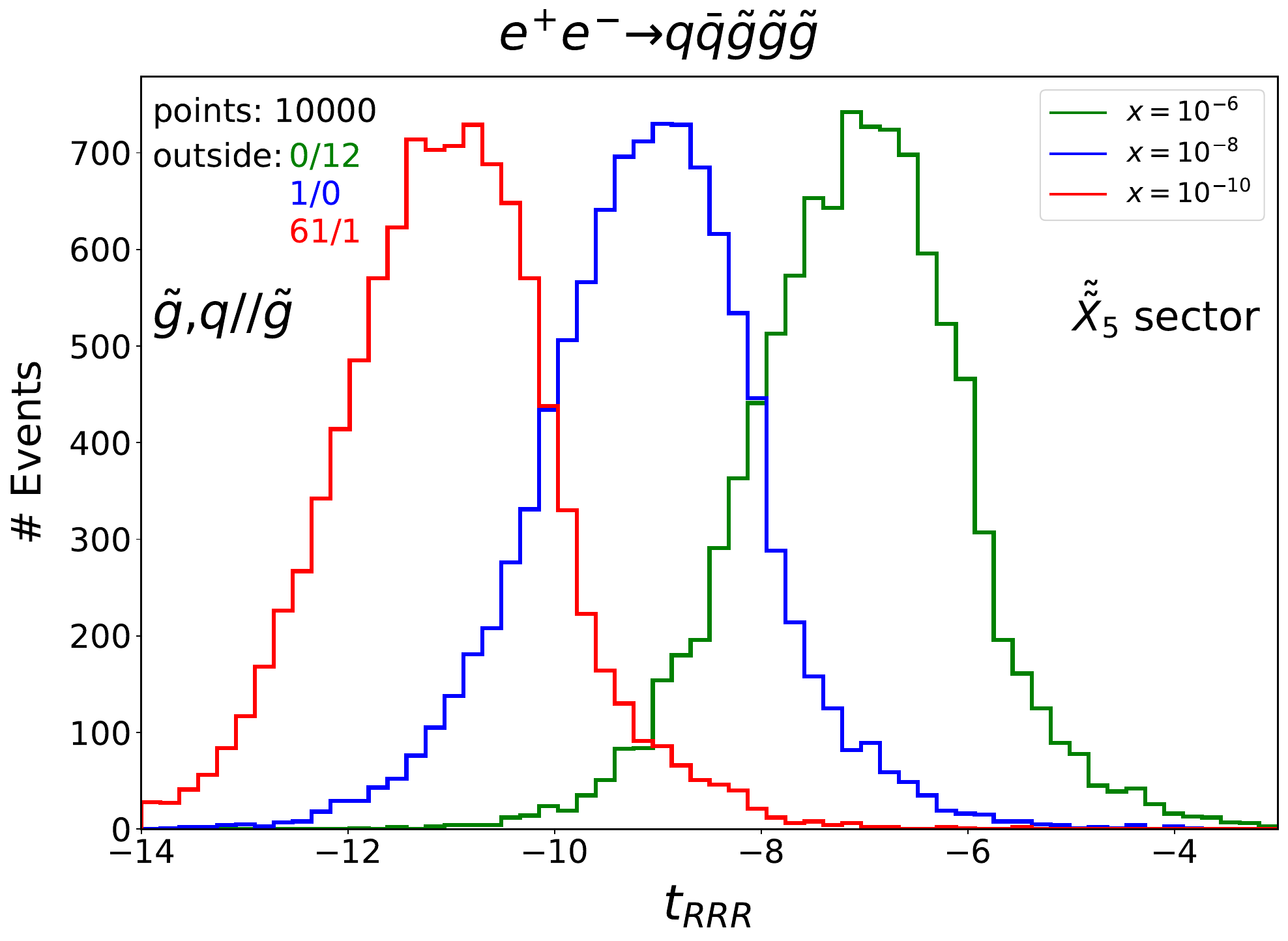}
	\includegraphics[width=0.32\columnwidth]{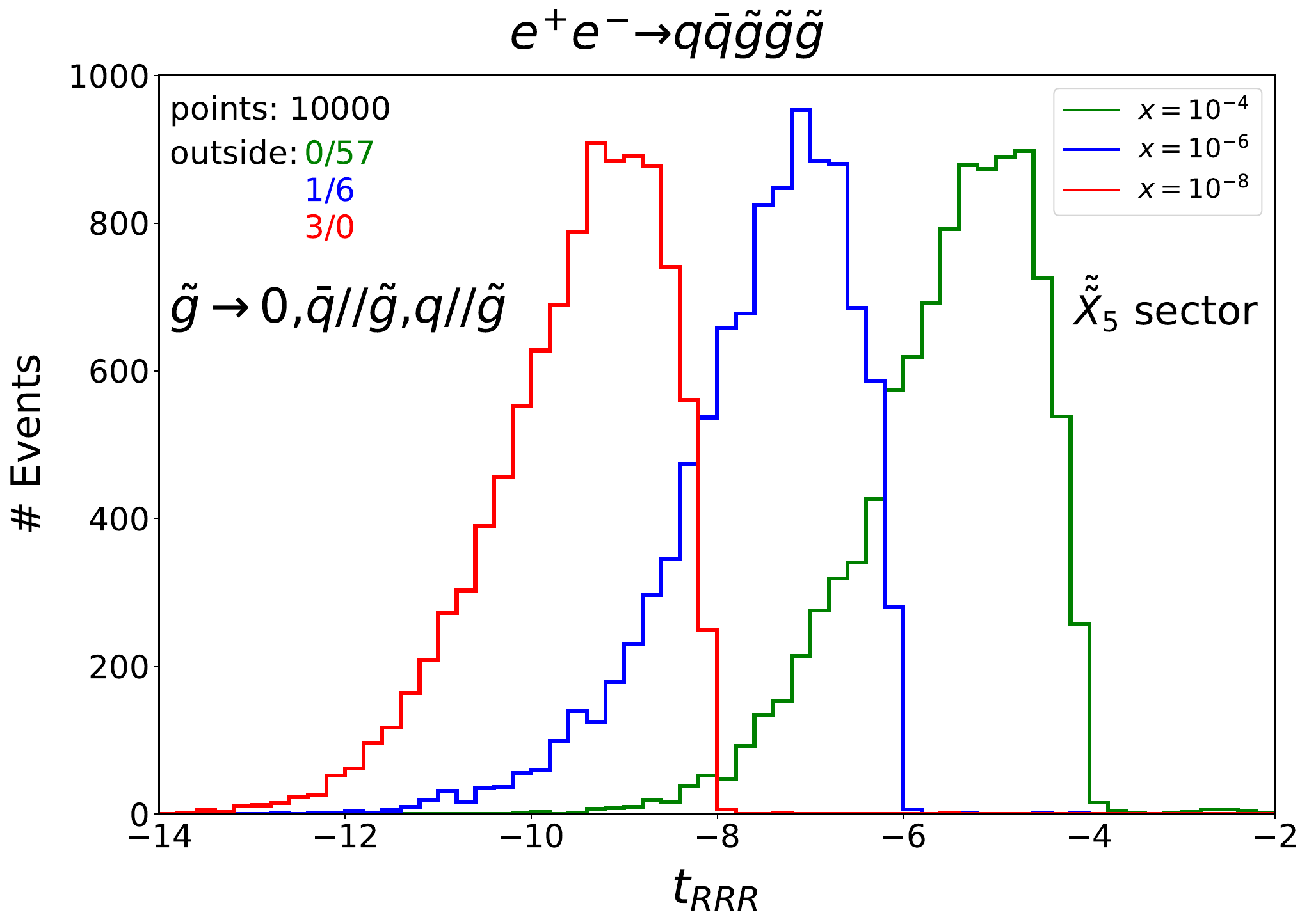}
	\includegraphics[width=0.32\columnwidth]{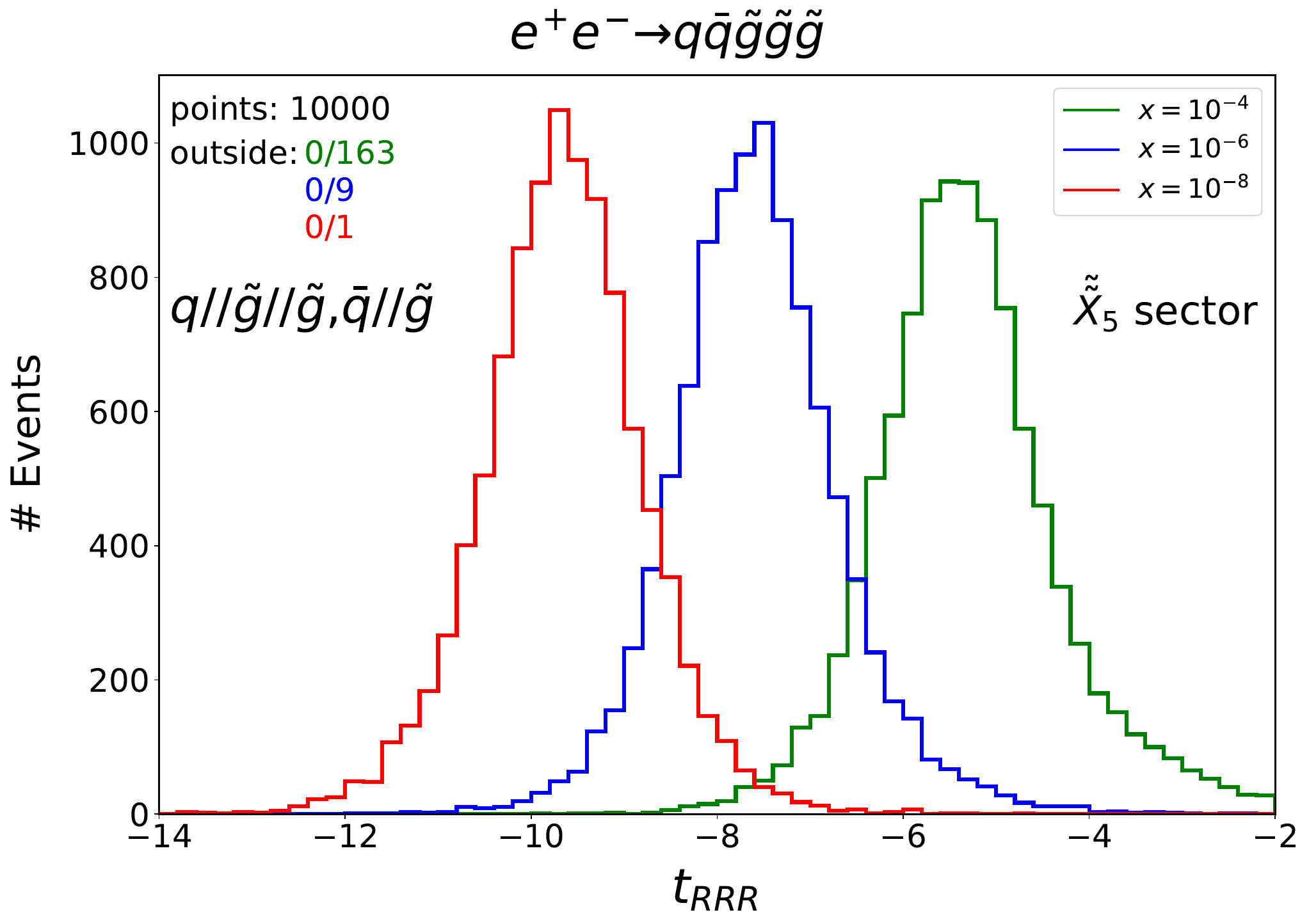}
	\caption{Distribution of the $t_{RRR}$ variable for the most subleading-colour contribution (all abelian gluons) to the $e^+ e^-\to q\bar{q}ggg$ sub-process, using the $\tilde{\tilde{A}}_5^0$ antenna function. The considered infrared limits are: single-collinear (upper left),  triple-collinear (upper centre), double-collinear (upper right), soft-single-collinear (lower left), soft-double-collinear (lower centre) and triple-collinear-single-collinear (lower right).}
	\label{fig:Att50}
\end{figure}

We present here a series of point-by-point tests for the cancellation of collinear singularities exploiting the definition of  triple-unresolved phase-space sectors for the assignment of momentum mappings in triple-real subtraction terms. In particular, we refer to the $\vardbtilde{A}_5^0$, $\wt{A}_5^0$ and $C_5^0$ antenna functions used for the N$^3$LO calculation in~\cite{Chen:2025kez}. These three antenna functions correspond to the three scenarios described in Section~\ref{sec:3unsector}.

\begin{figure}[hbt!]
 \centering
\includegraphics[width=0.32\columnwidth]{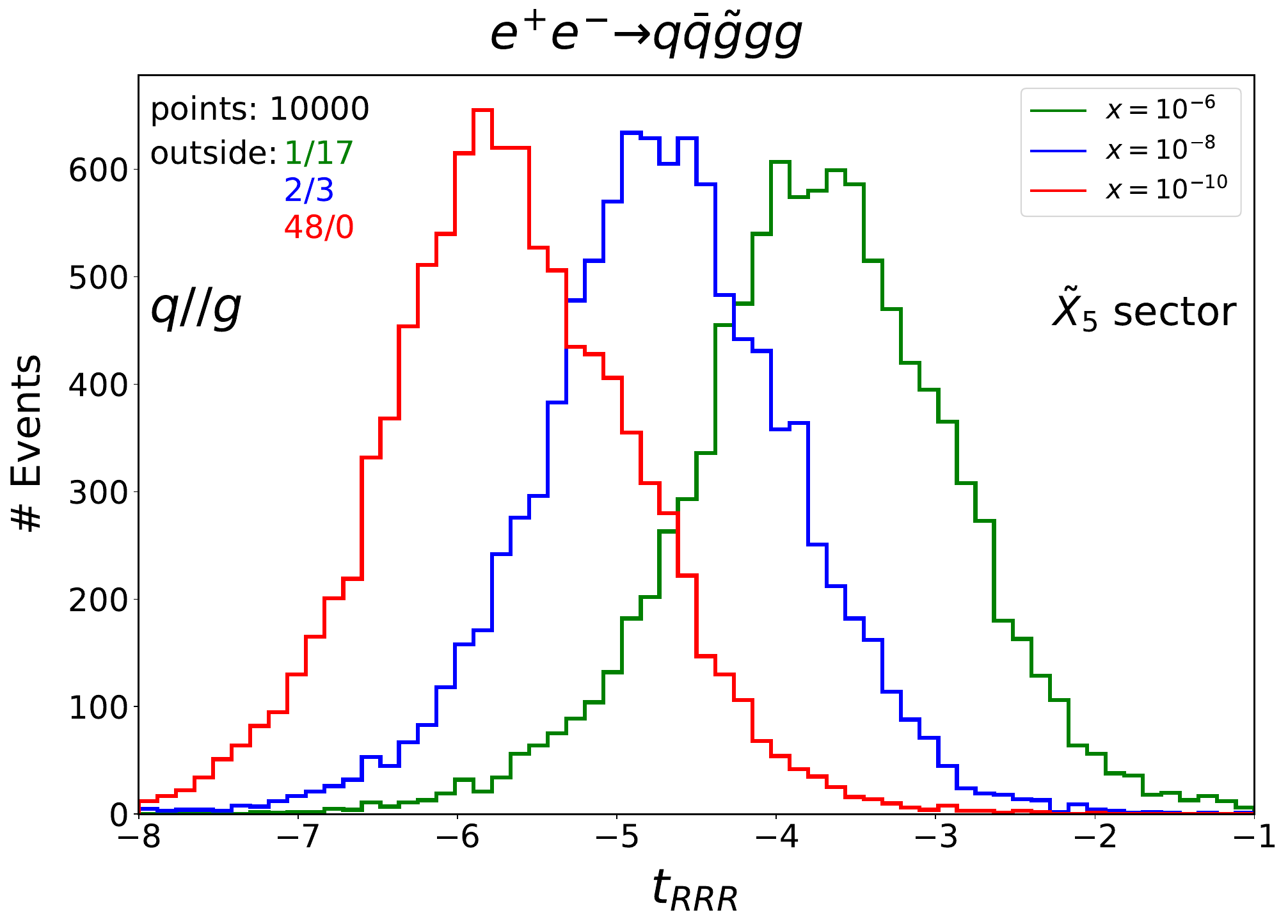}
\includegraphics[width=0.32\columnwidth]{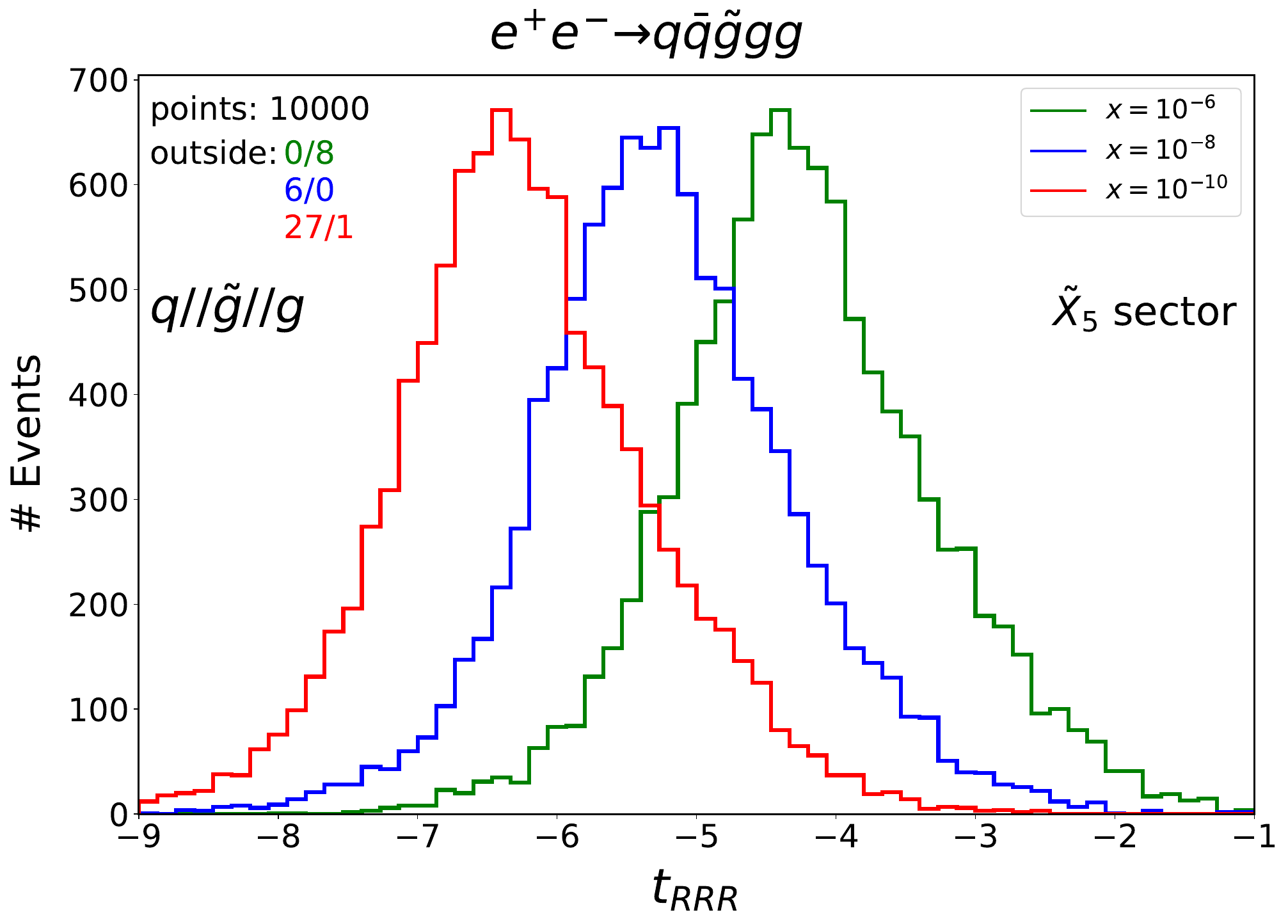}
\includegraphics[width=0.32\columnwidth]{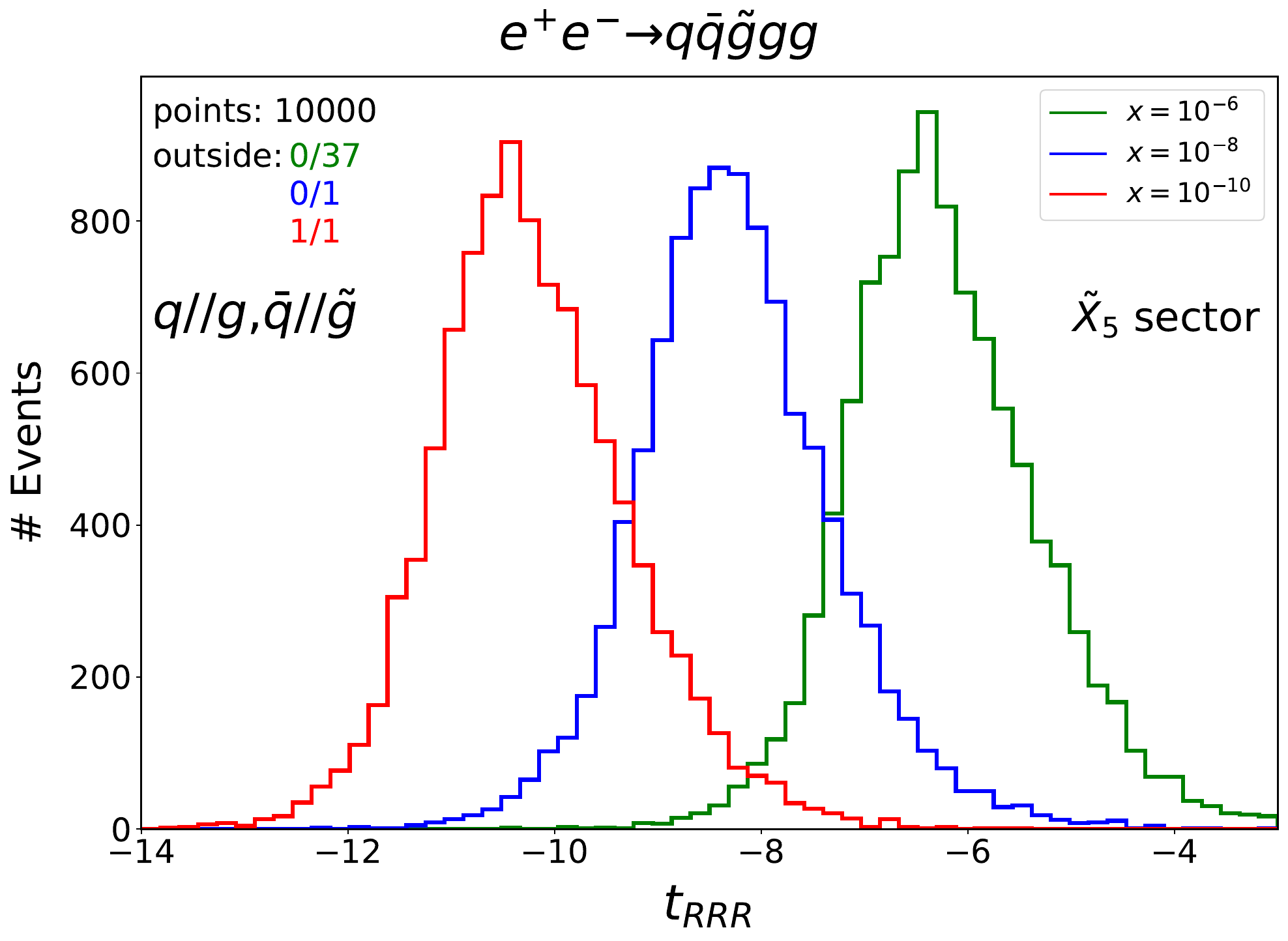}\\
\includegraphics[width=0.32\columnwidth]{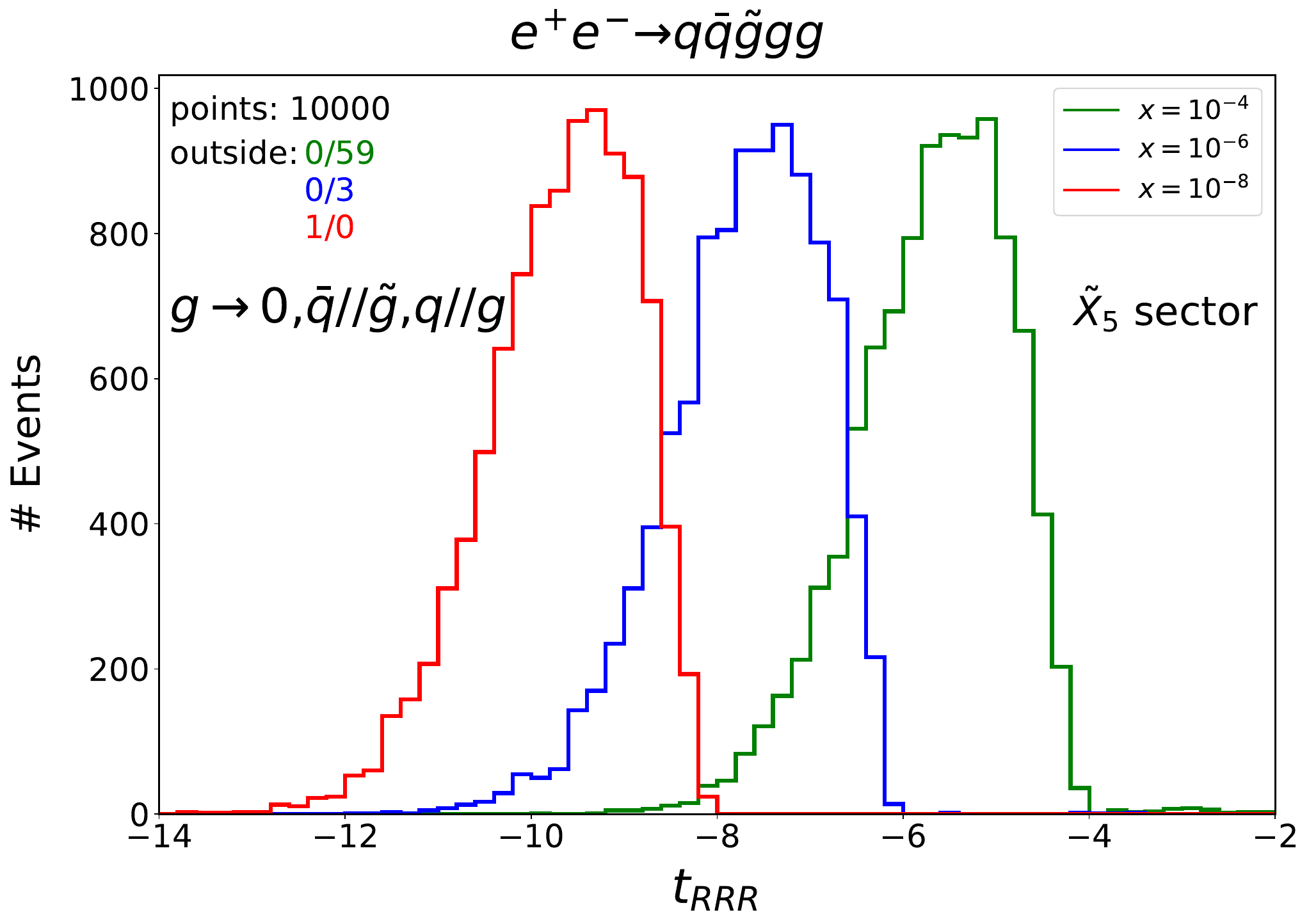}
\includegraphics[width=0.32\columnwidth]{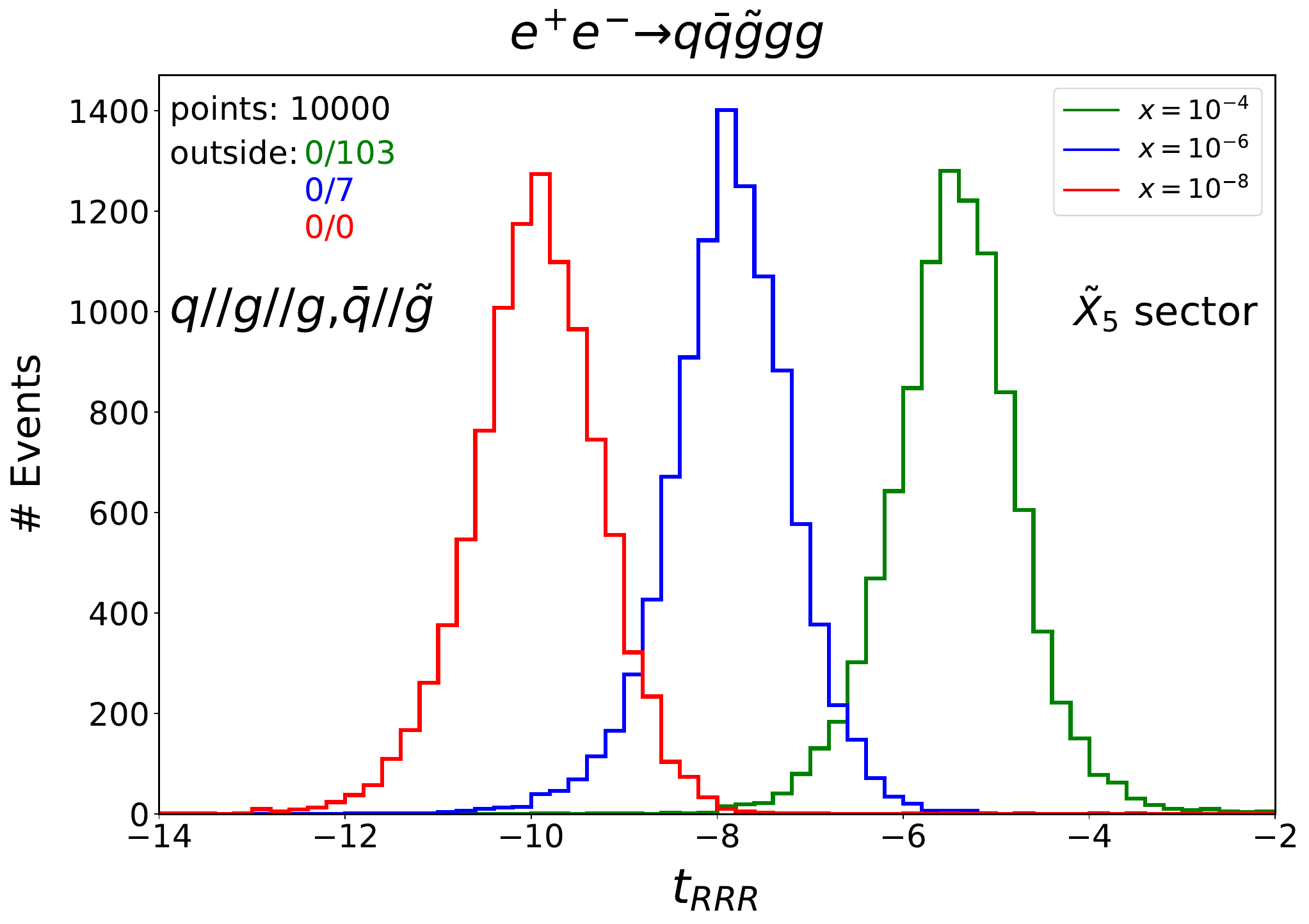}
\caption{Distribution of the $t_{RRR}$ variable for the subleading-colour contribution to the (one abelian gluon)  $e^+ e^-\to q\bar{q}ggg$ sub-process, using the $\wt{A}_5^0$ antenna function. The considered infrared limits are: single-collinear (upper left),  triple-collinear (upper centre), double-collinear (upper right), soft-double-collinear (lower left) and triple-collinear-single-collinear (lower right).}
\label{fig:At50}
\end{figure}

\begin{figure}[hbt!]
 \centering
\includegraphics[width=0.49\columnwidth]{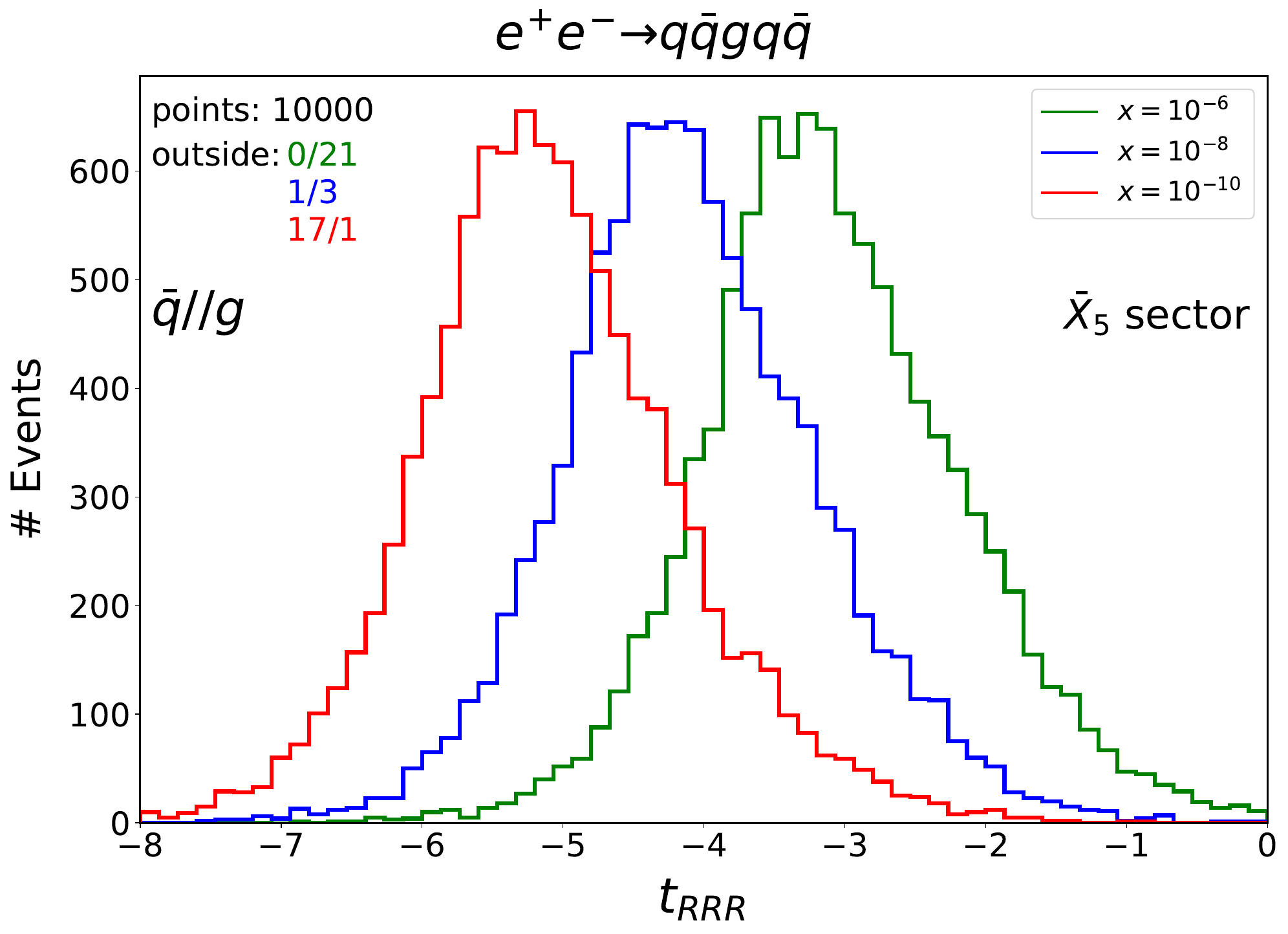}
\includegraphics[width=0.49\columnwidth]{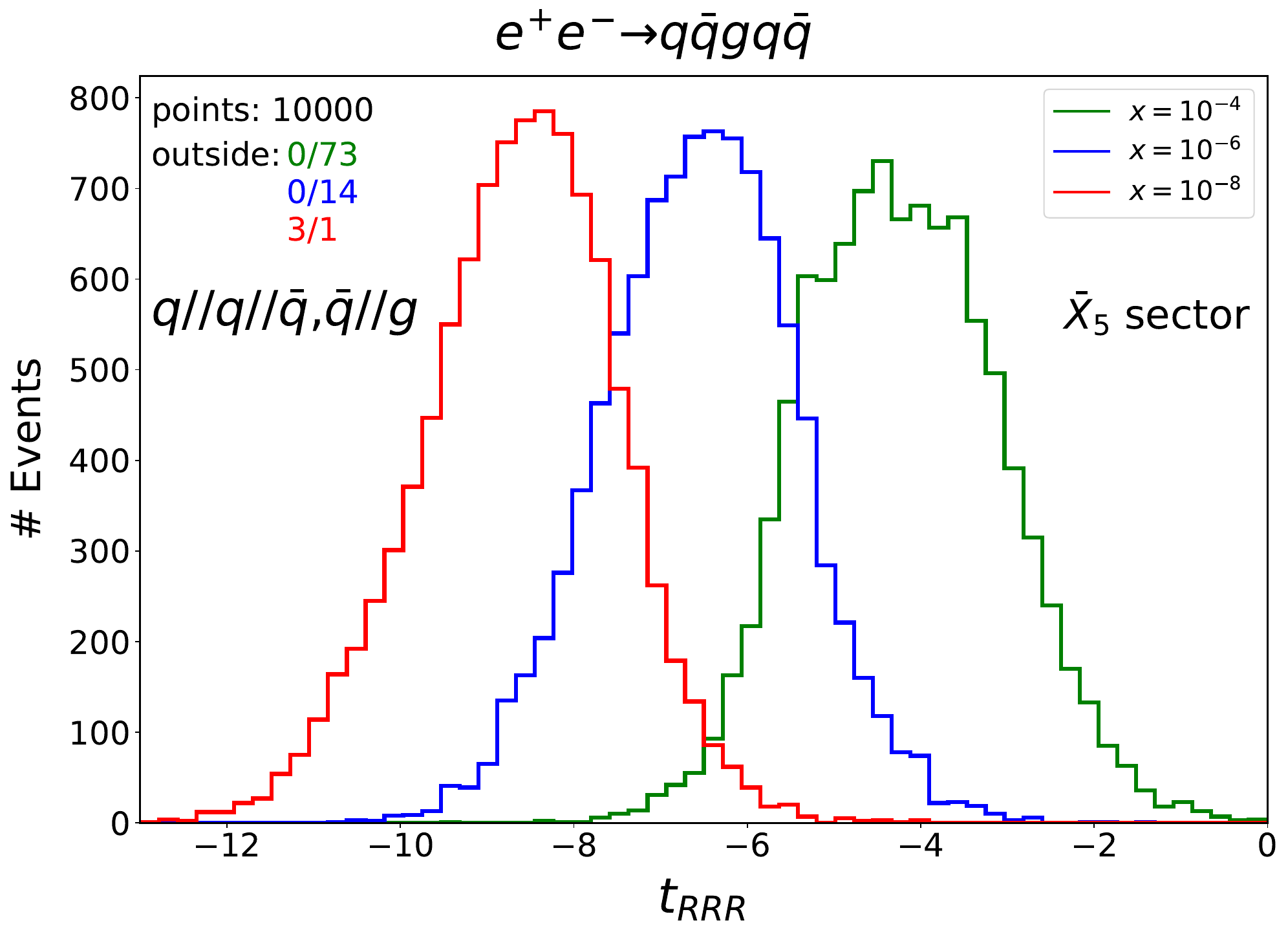}
\caption{Distribution of the $t_{RRR}$ variable for the same-flavour  $e^+ e^-\to q\bar{q}gq\bar{q}$ sub-process, using the $C_5^0$ antenna function. The considered infrared limits are: single-collinear (left),  and triple-collinear-single-collinear (right).}
\label{fig:C50}
\end{figure}

The quality of the cancellation between the matrix element and the subtraction term is assessed computing $t_{RRR}$, defined analogously to $t_{RR}$ in~\eqref{eq:tRR}. The results are shown in Figure~\ref{fig:Att50},~\ref{fig:At50} and~\ref{fig:C50} for the $\vardbtilde{A}_5^0$, $\wt{A}_5^0$ and $C_5^0$ antenna function respectively.  

We focus on all possible single-, double- and triple-unresolved configurations at N$^3$LO which can suffer from a mis-reconstructed pair of hard momenta mentioned in Section~\ref{subsubsec:3unresolvedmapping}. In all cases, we observe a consistent improvement of the infrared cancellation the deeper the collinear limits are probed, with a degree of convergence similar to the one observed at NNLO. We then conclude that the phase-space sectors are properly defined and the correct hard momenta are reconstructed in all scenarios. Finally, we note that the recovery of known results discussed in~\cite{Chen:2025kez} also stands as a solid proof of the correctness of our implementation.


\section{Conclusions}\label{sec:conclusions}

In this paper we address the problem of using ordered momentum mappings within unordered unresolved factors. This issue arises for example in the numerical implementation of the antenna subtraction method for higher-order calculations in QCD, where the antenna mapping is used to reabsorb unresolved final-state emissions onto the underlying hard configuration. 

We discuss how a simple decomposition of the phase space into sectors can resolve the ambiguity in the choice of momentum mapping. In particular, the sectors are defined via quadratic inequalities between Mandelstam invariants of the potentially unresolved emissions and the corresponding hard radiators to contain only a subset of the possible infrared limits which can be assigned to a specific ordering of momenta. Within each sector, a unique ordered mapping correctly reabsorbs the recoil of the additional emissions onto the hard radiators. We illustrate the definition of the relevant phase-space sectors up to three unresolved emissions (needed for N$^3$LO calculations) and provide analytical and numerical evidence of the correctness of the procedure. With respect to previous solutions to this problem, which consists in partial fractioning the unordered antenna functions into sub-antennae addressing only specific infrared divergences, the strategy proposed here does not require any antenna-specific work, hence is more general, easily scales at higher orders and prevents potential large cancellations in intermediate steps of the calculation.

The implementation discussed in this paper is particularly relevant for the extension of the antenna subtraction method to N$^3$LO and has already been applied for the first fully-differential calculation at this perturbative order with antenna subtraction~\cite{Chen:2025kez}. 

\section*{Acknowledgements}

We are grateful to Petr Jakub\v{c}\'{i}k and Giovanni Stagnitto for their contribution to the results presented in~\cite{Chen:2025kez}, where the implementation described in this paper was applied. We also thank Thomas Gehrmann and Nigel Glover for inspiring discussions and valuable feedback on the manuscript.
XC is supported by the National Science Foundation of China (NSFC) with grants No.12475085 and No.12321005. MM is supported by a Royal Society Newton International Fellowship (NIF/R1/232539).

\bibliographystyle{JHEP}
\bibliography{SectorMapping}

\end{document}